\documentclass[fleqn,usenatbib,usedcolumn]{mnras}
\pdfoutput=1

\usepackage[british]{babel}             

\usepackage{graphicx}
\usepackage{natbib}
\usepackage{amsmath}
\usepackage{verbatim}
\usepackage{color}
\usepackage[dvipsnames]{xcolor}
\usepackage{mathtools}
\usepackage{tikz}
\usepackage{multirow}
\usepackage{caption}
\usepackage{amsmath,esint} 

\usepackage{newtxtext,newtxmath}
\usepackage{amssymb}    
\usepackage{wasysym}
\usepackage[T1]{fontenc}
\usepackage{ae,aecompl} 

\usepackage{mathtools, cuted}
\usepackage{lipsum, color}
\usepackage[export]{adjustbox}
\usepackage{blindtext}

\usetikzlibrary{shapes.geometric, arrows,positioning}
\usetikzlibrary{calc}
\tikzstyle{box} = [rectangle, rounded corners, minimum width=4cm, minimum height=1cm, text centered, text width=4cm, draw=black, fill=white!30]

\tikzstyle{boxRK} = [rectangle, rounded corners, minimum width=2cm, minimum height=2cm, text centered, text width=2cm, draw=black, fill=white!30]

\tikzstyle{loop} = [diamond, minimum width=4cm, aspect=2.5, text centered, draw=black, fill=green!30]

\tikzstyle{stextr} = [rectangle, minimum width=4cm, text centered, draw=red, text=red, fill=white,opacity=0,text opacity=1]

\tikzstyle{stext} = [rectangle, minimum width=4cm, text centered, draw=black, fill=white,opacity=0,text opacity=1]

\tikzstyle{arrow} = [thick,->,>=stealth]

\newcommand{\mar}{2cm}
\newcommand{\marboxx}{0.15}
\newcommand{\marboxy}{0.15}
\newcommand{\mardox}{0.3}
\newcommand{\mardoy}{0.2}
\newcommand{\marnextx}{-0.4}
\newcommand{\marnexty}{-0.3}
\newcommand{\mardonex}{0.4}
\newcommand{\mardoney}{-0.5}
\newcommand{\pat}{\partial_t}
\newcommand{\vnabla}{{\bf \nabla}}

\newcommand{\vx}{{\bf x}}
\newcommand{\vy}{{\bf y}}
\newcommand{\vz}{{\bf z}}
\newcommand{\vb}{{\bf B}}
\newcommand{\vv}{{\bf v}}
\newcommand{\vE}{{\bf E}}
\newcommand{\vf}{{\bf f^x}}
\newcommand{\vg}{{\bf f^y}}
\newcommand{\vh}{{\bf f^z}}
\newcommand{\Ee}{\overline{E}}
\newcommand{\bu}{\overline{\bf U}}
\newcommand{\Ef}{{{\Phi}}}

\newcommand{\facex}{S_x}
\newcommand{\facey}{S_y}
\newcommand{\facez}{S_z}

\newcommand{\vF}{{\bf \overline{F}^x}}
\newcommand{\vG}{{\bf \overline{F}^y}}
\newcommand{\vH}{{\bf \overline{F}^z}}

\newcommand{\alfven}{Alfv\'{e}n\ }

\newcommand{\vU}{{\bf U}}

\newcommand{\Dt}{\Delta t}
\newcommand{\Dx}{\Delta x}
\newcommand{\Dy}{\Delta y}
\newcommand{\Dz}{\Delta z}
\newcommand{\beqa}{\begin{eqnarray}}
\newcommand{\eeqa}{\end{eqnarray}}
\newcommand{\beq}{\begin{equation}}
\newcommand{\eeq}{\end{equation}}

   \volume{{\rm in press}}

\title[]{Fourth Order Accurate Finite Volume CWENO Scheme For Astrophysical MHD Problems}
 \author[Verma et al.]{Prabal Singh Verma$^{1,2,4}$\thanks{E-mail:prabal-singh.verma@univ-amu.fr},
    Jean-Mathieu Teissier$^{1,3}$,
    Oliver Henze$^{1}$,
    \newauthor
    Wolf-Christian M\"uller$^{1,2,3}$
\\
$^{1}$  Technische Universit\"at Berlin, ER 3-2, Hardenbergstr. 36a, 10623 
	Berlin, Germany\\
$^{2}$ Max-Planck/Princeton Center for Plasma Physics, Princeton, NJ, USA  \\
$^{3}$ Berlin International Graduate School in Model and Simulation Based Research \\
$^{4}$ CNRS, Aix-Marseille Univ., PIIM, UMR 7345, Marseille F-13397, France
}

\date{Accepted XXX. Received YYY; in original form ZZZ}

\pubyear{2018}

\begin{document}
\label{firstpage} 
\pagerange{\pageref{firstpage}--\pageref{lastpage}}
\maketitle

\begin{abstract}
In this work, a simple
fourth-order accurate finite volume semi-discrete scheme is introduced to solve astrophysical magnetohydrodynamics (MHD) problems on Cartesian meshes. Hydrodynamic quantities like density, momentum and energy are discretised as volume averages. The magnetic field and electric field components are discretised as area and line averages respectively, so as to employ the constrained transport technique, which preserves the solenoidality of the magnetic field
to machine precision. The present method makes use of a dimension-by-dimension approach employing a 1-D fourth-order accurate centrally weighted essentially non-oscillatory (1D-CWENO4) reconstruction polynomial. A fourth-order accurate, strong stability preserving (SSP) Runge-Kutta method is used to evolve the semi-discrete MHD equations in time. Higher-order accuracy of the scheme is confirmed in various linear and nonlinear multi-dimensional tests and the robustness of the method
in avoiding unphysical numerical artifacts in the solution 
 is demonstrated through several complex MHD problems.
        \end{abstract}

        \maketitle
        \textsl{\textsl{}}

\section{Introduction}
Many astrophysical phenomena can be approximately described by the ideal magnetohydrodynamics (MHD) equations.
Since astrophysical flows are often highly-supersonic, discontinuities and shocks are very frequent. This is the reason why various shock-capturing numerical methods have been proposed to solve astrophysical MHD problems. Many of them are based on finite differences, where point values are evolved in time, e.g., \citep{van1979towards,EVH88,1995ApJ...442..228R}, or on the finite volume method, where cell averages are evolved in time \citep{godunov1959difference,zachary1992higher,zachary1994higher,dai1998simple,balsara1999maintaining,
balsara1999staggered,janhunen2000positive,toth2000b,dedner2002hyperbolic,balsara2004second,ziegler2004central}. Usually, the existing MHD codes, based on the above-mentioned algorithms, are second-order accurate. Efforts to improve the quality of numerical solutions focus on the reduction of numerical dissipation. To this end, second-order schemes can be combined with computationally expensive Riemann solvers in order to obtain less diffusive numerical results \citep{roe1981approximate,toro1994restoration,powell1997approximate,toro2013riemann}.

Another way to reduce the numerical dissipation is to use high-order schemes. These schemes have received significant attention for both finite differences, e.g., \citep{shu1988efficient,shu1989efficient,delzanna:ctdivb,mignone2010high,toth2014high} and finite volume frameworks, e.g., \citep{woodward1984numerical,colella1984piecewise,shu2009high}.
In both methodologies, higher-order accuracy in smooth regions is achieved through higher-order reconstruction polynomials.
Therefore, codes based on higher-order schemes, without the use of expensive Riemann solvers \citep{lpr4,cs12,BUH14,ROM15,verma2017higher}, exhibit much lower numerical dissipation as compared to second-order schemes. However, codes based on second-order methods \citep{stone1992zeus,ziegler2004central} have the advantage of being more robust against the generation of numerical artifacts in the solution than higher-order schemes when dealing with flows with discontinuities and shocks. In addition, they can more easily be combined with adaptive mesh refinement (AMR) techniques, e.g. \citep{berger1984adaptive, berger1989local, grauer1998adaptive, balsara2001divergence, tang2003adaptive, anderson2006relativistic, zhang2006ram, rosenberg2007adaptive, van2007hybrid, hu2013adaptive, dumbser2013ader,buchmueller_dreher_helzel:colellaamr}.
Therefore, the present work aims to suggest an efficient and simple higher-order finite volume scheme for MHD flows.
As mentioned above, higher-order schemes require higher-order shock capturing polynomials. The most popular
ones are the weighted essentially non-oscillatory (WENO) polynomials \citep{liu1994weighted,jiang1996efficient} which are based on the ENO approach \citep{harten1987uniformly}.
Recently, very high order WENO schemes from the $7th$ to the $17th$ order of accuracy have been suggested for hydrodynamic
flows \citep{weno1,weno4,weno5,shu2009high,gerolymos2009very,BUH14}.

In the present work, we choose to employ 1-D fourth-order accurate centrally weighted essentially non-oscillatory (1D-CWENO4) reconstruction polynomials \citep{lpr7} in the dimension-by-dimension framework
(as the one in \cite{verma2017higher}). The first central scheme was proposed in \cite{lxf}, which employs the spatial averaging of neighbouring grid cells as part of the integration step.
This procedure can be considered as an imprecise and highly diffusive partial approximation of a Riemann problem.
The Lax-Friedrichs scheme, being first-order accurate, turns out to be numerically too dissipative to be of practical use.
Therefore, in order to reduce the numerical viscosity, a second-order central scheme based on a non-oscillatory reconstruction of the linear interpolant has been developed \citep{nt,jiang1998nonoscillatory}. This approach was further improved by introducing second and third-order semi-discrete central schemes, e.g. \citep{kurganov-2000,kt-2000,kp-2001,knp-2001,kl-2002,kt-2002}. The heart of these semi-discrete central schemes is a CWENO reconstruction of the local polynomial under consideration. Numerous third and fourth-order CWENO reconstruction methods have been
suggested for 1D, 2D and 3D hyperbolic conservation laws \citep{lpr6,lpr7,lpr5,lpr4,cs12,cs15}. The CWENO
method has also been employed for nonuniform meshes and for AMR \citep{cweno-2016,cweno-weno-2016}.

Note that since we are proposing a semi-discrete scheme here, one may alternatively apply other well-known
higher-order non-oscillatory reconstructions, e.g., PPM limiter \citep{ppm}, CWENO5 \citep{capdeville2008central},
WENO \citep{shu2009high}, MP5 \citep{suresh1997accurate}, etc. When solving the MHD equations, a major difficulty one encounters is that $\vnabla \cdot \vb$ usually grows with time, if one treats the magnetic field components like the other hydrodynamic variables. Thus, in MHD simulations, a crucial issue is to maintain the solenoidality of the magnetic field ($\vnabla \cdot \vb=0$) as the system evolves in time. Otherwise, non-physical effects arise \citep{brackbill1980effect,powell1994approximate}.
Several methods have been proposed in order to solve this issue. Frequently, a {\it divergence-cleaning} approach is used
where the numerical errors associated with $\vnabla \cdot \vb$ are removed by employing one of the following three widely used techniques, namely, the {\it Helmholtz-Hodge projection} \citep{brackbill1980effect}, the {\it 8-wave Multiplier method} \citep{powell1994approximate} and the {\it Generalized Lagrange Multiplier technique} \citep{dedner2002hyperbolic}. The projection method creates spurious small-scale fluctuations
\citep{balsara2004comparison} as a consequence of the divergence-free but physically not fully consistent state of the projected magnetic field. The 8-wave method is
neither directly avoiding nor eliminating magnetic monopoles. Instead, the deviations from solenoidality are advected by the flow velocity.
The scheme by Dedner et al. couples the evolution of the magnetic field to the solenoidality constraint through an additional scalar field \citep{dedner2002hyperbolic} and
has been used to develop a high-order dimension-by-dimension WENO scheme to simulate both non-relativistic \citep{ROM15} and relativistic MHD flows \citep{ROM16}.
However, in the present work, we make use of another classical approach, the {\it constrained transport} (CT) technique \citep{EVH88} which preserves the solenoidality of the magnetic
field up to machine precision.

The CT technique combined with second-order schemes is being used extensively to study astrophysical MHD
problems, e.g., \citep{balsara2004second,ziegler2004central,gardiner2005unsplit,gardiner2008unsplit,rossmanith2006unstaggered,
helzel2011unstaggered}. Although third-order accurate CT-finite volume methods have been suggested in \citep{LOZ00,LOZ04}, we emphasise here that it is not straightforward to use the CT technique
to develop a fourth-order finite volume scheme in a way that yields higher-order accuracy in smooth regions and at the same time stays robust in the vicinity of strong MHD shocks.
In this approach, hydrodynamic quantities (density, momentum and energy) are defined as volume averages, whereas
the magnetic field and electric field components
are defined as area averages and line averages, respectively.
Thus, the presence of three simultaneous discretisations introduces complexity in computing higher-order area-averaged numerical fluxes (discussed later in the manuscript) and line-averaged electric field components. A precise computation of these quantities is essential to maintain higher-order accuracy and to limit the appearance of numerical artifacts over time. This paper aims at giving a simple and efficient fourth-order accurate finite volume CWENO scheme that uses
the CT technique. Contrary to other higher-order methods \citep{balbas2006nonoscillatory,li2010fourth,ROM15}, where
several point values have to be reconstructed to estimate the numerical fluxes, the method presented in this paper needs only two reconstructions per numerical mesh cell per spatial dimension. The higher-order accuracy in smooth regions is maintained through area-average$\leftrightarrow$point value
transformations \citep{ppm,BUH14,verma2017higher}.

This work is organised as follows. In section \ref{sec:equations}, we formulate the ideal MHD equations to be solved as well as the finite volume approach. Then, we explain step-by-step how the right-hand sides of the hydrodynamical equations
are computed in section \ref{sec:RHS}. Section \ref{sec:CT} tells how the magnetic field is evolved in time while preserving its
solenoidality using the CT method. A simple method enhancing the robustness of the numerical method by reducing the appearance of unphysical numerical fluctuations is presented in section \ref{sec:negative}. The time integration and a flow-chart of the method are described in section \ref{sec:timeINT}. The validity of the numerical method and its advantage with respect to a lower-order scheme have been confirmed through various multidimensional numerical tests, presented in section \ref{sec:tests}.  Finally, section \ref{sec:conclusion} gives some concluding remarks.
\vspace{-0.3in}
\section{Ideal MHD equations and numerical framework} \label{sec:equations}
\subsection{Equations of Ideal Magnetohydrodynamics}
The ideal MHD equations, in conservative form, can be expressed as follows:
        \beqa
         \label{eq:mhdrho}
         \pat \rho  &=& -\vnabla \cdot (\rho \vv),
        \eeqa 
        \beqa
         \pat (\rho \vv) &=& - \vnabla \cdot \left( \rho \vv \vv^T + (p + \frac{1}{2}|\vb|^2)I - \vb \vb^T \right),
        \eeqa 
        \beqa
         \label{eq:mhde}
         \pat  e  &=& -\vnabla \cdot \left((e + p + \frac{1}{2}|\vb|^2)\vv  -  (\vv\cdot\vb)\vb \right),
        \eeqa
        \beqa
         \label{eq:mhdb}
         \pat \vb &=& -\vnabla \times \vE,
        \eeqa
         with the constraint
         \beqa \label{divb}
                \vnabla \cdot \vb &=& 0.
         \eeqa
        Here $I$ is the $3\times 3$ identity matrix, $\rho$ the mass density, $\vv$ the velocity, $\vb$ the magnetic field, $e$
        the total energy density and $\vE = - \vv \times \vb$ the electric field.
        The thermal pressure $p$ is computed from the ideal gas equation of state, assuming that the
thermodynamic changes of state
are adiabatic:
         \beqa \label{eq:pressure}
          p &=&  (\gamma -1) \left(e - \frac{1}{2}\rho |\vv|^2 - \frac{1}{2}|\vb|^2 \right),
         \eeqa
        where $\gamma$ is the ratio of specific heats.
        \subsection{Finite Volume Approach and Formalism}
In the finite volume discretisations, the hydrodynamic variables $(\rho,\rho v_x,\rho v_y,\rho v_z,e)$ $\equiv {\bf U}$
are defined as volume-averaged quantities,
denoted by $\bu_{i,j,k}$. The magnetic field components $(B_x, B_y, B_z)$ $\equiv {\bf B}$ are defined as
area averages $\overline{\bf B}_{i,j,k}^F$.
Here the subscripts $(i,j,k)$ represent the grid-cell under consideration.
The volume-averaged quantity, $\bu_{i,j,k}$ can be computed by integrating the physical quantity ${\bf U}$ over
the region of space
$\Omega_{i,j,k}=[x_i-\frac{\Dx}{2},x_i+\frac{\Dx}{2}]\times[y_j-\frac{\Dy}{2},y_j+\frac{\Dy}{2}]\times[z_k-\frac{\Dz}{2},z_k+\frac{\Dz}{2}]$.
Here $x_i=(i+0.5)\Dx$, $y_j=(j+0.5)\Dy$, $z_k=(k+0.5)\Dz$, and $\Dx,\Dy,\Dz$ are
grid-sizes in
the $\vx$-, $\vy$- and $\vz$-directions, respectively.
Mathematically, we can express $\bu_{i,j,k}$ as,
\beqa
\label{u_avg}
\bu_{i,j,k}&=&\frac{1}{\Dx\Dy\Dz}\int_{\Omega_{i,j,k}} {\mathrm{d}x  \, \mathrm{d}y \, \mathrm{d}z \,} \vU(x,y,z).
\eeqa
Integrating the equations \eqref{eq:mhdrho}-\eqref{eq:mhde} over $\Omega_{i,j,k}$ and
applying Gauss' theorem, we end up with semi-discrete equations in the following form,
\begin{figure}
\includegraphics[width=0.75\columnwidth,center]{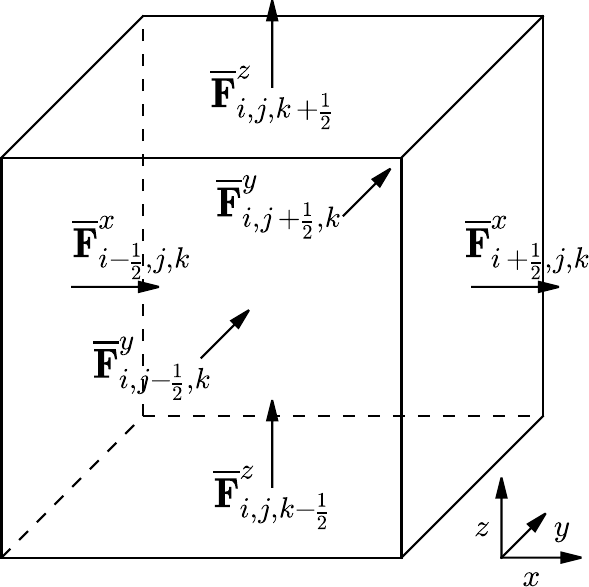}
\caption{ \label{fig:fluxesFGH}Definition of the fluxes $\vF$, $\vG$ and $\vH$.}
\end{figure}
\beqa
\label{eq:compactU}
\nonumber \frac {d\bu_{i,j,k}}{dt} &=&
-\frac{\vF_{i+1/2,j,k}-\vF_{i-1/2,j,k}}{\Dx}  \\ \nonumber &&- \frac{\vG_{i,j+1/2,k}-\vG_{i,j-1/2,k}}{\Dy}
\\ &&- \frac{\vH_{i,j,k+1/2}-\vH_{i,j,k-1/2}}{\Dz},
\eeqa
where 
$\vF$, $\vG$ and $\vH$ are the area-averaged fluxes in the $\vx$-, $\vy$- and $\vz$-directions, respectively,
as shown in figure \ref{fig:fluxesFGH}. These fluxes are expressed as,
\beqa
\label{eq:fluxF}
\vF_{i\pm1/2,j,k}&=&\frac{1}{\Dy\Dz}\int_{\facex}  {\mathrm{d}y \, \mathrm{d}z \, \vf(x_{i\pm1/2},y,z)},\\
\label{eq:fluxG}
{\vG_{i,j\pm1/2,k}}&=&\frac{1}{\Dx\Dz}\int_{\facey} {\mathrm{d}x  \, \mathrm{d}z \, \vg(x,y_{j\pm1/2},z)},\\
\label{eq:fluxH}
{\vH_{i,j,k\pm1/2}}&=&\frac{1}{\Dx\Dy}\int_{\facez} {\mathrm{d}x  \, \mathrm{d}y \, \vh(x,y,z_{k\pm1/2})},
\eeqa
with $\facex = [y_j-\frac{\Dy}{2},y_j+\frac{\Dy}{2}]\times[z_k-\frac{\Dz}{2},z_k+\frac{\Dz}{2}]$. $\facey$
and $\facez$ are defined analogously.
The quantities $\vf$, $\vg$ and $\vh$ are the point-valued fluxes, which are described below as,
\beqa
\label{eq:fluxf}
{\vf} &=& {\vf}{\bf(U, B)} =\begin{pmatrix} \rho v_x \\ \rho v_x^2+p+|\vb|^2/2-B_x^2 \\ \rho v_xv_y-B_xB_y \\ \rho v_xv_z-B_xB_z \\ (e+p+|\vb|^2/2)v_x-B_x(\vv \cdot \vb) \end{pmatrix},\\
\label{eq:fluxg}
{\vg} &=& {\vg}{\bf(U, B)} =\begin{pmatrix} \rho v_y \\ \rho v_xv_y-B_xB_y \\ \rho v_y^2+p+|\vb|^2/2-B_y^2 \\ \rho v_yv_z-B_yB_z \\ (e+p+|\vb|^2/2)v_y-B_y(\vv \cdot \vb) \end{pmatrix},\\
\label{eq:fluxh}
{\vh} &=& {\vh}{\bf(U, B)} =\begin{pmatrix} \rho v_z \\ \rho v_xv_z-B_xB_z \\ \rho v_yv_z-B_yB_z \\ \rho v_z^2+p+|\vb|^2/2-B_z^2 \\ (e+p+|\vb|^2/2)v_z-B_z(\vv \cdot \vb) \end{pmatrix}.
\eeqa
We note here that the accuracy of the semi-discrete scheme in Eq.\eqref{eq:compactU} depends on the following:
\begin{enumerate}
\item Accuracy of the integrator employed to evolve the equations in time (discussed in section \ref{sec:timeINT}).
\item Accuracy of the area-averaged fluxes ($\vF$, $\vG$, $\vH$) at the grid-cell interfaces (discussed below as well as in section \ref{sec:RHS} in detail).
\end{enumerate}
The latter plays a major role in the numerical dissipation of a semi-discrete scheme.
However, we learn from Eqs.\eqref{eq:fluxF}-\eqref{eq:fluxH} that the accuracy of the area-averaged fluxes itself depends
on the accuracy of the point-valued fluxes ($\vf$, $\vg$ $\vh$).
Since these point-valued fluxes are nonlinear functions
of the physical quantities, they can have an accuracy higher than second-order only if the point values of the
physical quantities at the grid-cell interfaces are higher-order accurate. In section \ref{sec:RHS} we will discuss how to
reconstruct fourth-order accurate point values of the physical quantities at the grid cell interface from the given volume averages and
then how to utilise these point values to compute fourth-order accurate area-averaged fluxes.

The CT procedure used to ensure the solenoidality of the magnetic field is described in section \ref{sec:CT}.
\section{Fluxes computation}
\label{sec:RHS}
The fluxes appearing
on the right-hand side of equation \eqref{eq:compactU} are computed for each dimension separately following a dimension-by-dimension approach,
e.g. \citep{weno5,shu2009high,BUH14,verma2017higher} through the following steps.
\begin{enumerate}
\item Reconstruct a 1D higher-order polynomial from the given volume averages.
\item Employ the polynomial to compute area-averaged quantities at the
grid-cell interfaces.
\item Deduce point values of the physical quantities at the face centres from the area averages.
\item Compute point-valued fluxes from already computed point values of the physical quantities.
\item Deduce area-averaged fluxes by integrating over point-valued fluxes.
\end{enumerate}
The next subsections describe each of these steps.
In the following, the computation of the flux in the $\vx$-direction is illustrated, that is, the $\vF$ terms in equation \eqref{eq:compactU}. The computations of the two other fluxes $\vG$ and $\vH$ are done in an analogous manner.
\subsection{Step 1: Reconstruction of a 1D Higher-Order Polynomial from Volume Averages}
\label{sec:reconstruction}
We construct here a 1D fourth-order CWENO polynomial from the given volume averages. Before going into the details of the procedure, we note here that the method for reconstructing such polynomials is well described in \cite{lpr3}. Nevertheless, we provide here a brief overview of the method for the sake of completeness. In this approach we reconstruct a quadratic polynomial ${\bf R}_{i,j,k}(x)$ in each cell $(i,j,k)$, which is a convex combination of three quadratic polynomials ${\bf P^i}_{i-1,j,k}(x)$, ${\bf P^i}_{i,j,k}(x)$ and ${\bf P^i}_{i+1,j,k}(x)$ such that,
\begin{equation} \label{R_I_4th}
{\bf R}_{i,j,k}(x) = \sum_{m = i-1}^{i+1} w^i_{m,j,k} {\bf P^i}_{m,j,k}(x).
\end{equation}
The quantities $w^i_{m,j,k}$ are the nonlinear weights which ensure higher-order accuracy in the smooth regions and non-oscillatory
behaviour near a discontinuity. These weights satisfy the following criteria,
\begin{equation} \label{weights_sum}
\sum_{m = i-1}^{i+1} w^i_{m,j,k} = 1
 ,\hspace{0.2cm} w^i_{m,j,k} \geq 0, \hspace{0.2cm}  \forall \hspace{0.1cm} m \in \{i-1, i, i+1\},
\end{equation}
and are defined as,
\begin{equation} \label{w_l}
w^i_{m,j,k} = \frac{\alpha^i_{m,j,k}}{\alpha^i_{i-1,j,k}+\alpha^i_{i,j,k}+\alpha^i_{i+1,j,k}},
\end{equation}
with, 
\begin{equation} \label{alpha}
\alpha^i_{m,j,k} = \frac{c_{m,j,k}}{(\epsilon + IS^i_{m,j,k})^p}, \hspace{0.2cm} \forall \hspace{0.1cm}m \in \{i-1, i, i+1\}.
\end{equation}
Here $\epsilon$, $p$ are chosen to be $10^{-6}$ and $2$, respectively and the constants $c_{i-1,j,k} = c_{i+1,j,k} = 1/6$, $c_{i,j,k} = 2/3$
are chosen so as to guarantee the fourth-order accuracy of the physical quantities at the cell-boundaries \citep{lpr3}.
The quantity $IS^i_{m,j,k}$ is the smoothness indicator quantifying the smoothness of the corresponding
polynomial ${\bf P^i}_{m,j,k}(x)$. It is defined as,
 \begin{eqnarray}\label{ISn}
\nonumber  IS^i_{m,j,k} &=& \sum_{n=1}^2 \int_{x_{i-1/2}}^{x_{i+1/2}} \! (\Delta x)^{2n-1}  ({\bf P^i}_{m,j,k}^{(n)}(x))^2 \,  \mathrm{d}x , 
 \end{eqnarray}
where $m \in \{i-1, i, i+1\}$ and $n$ represents the order of the derivative w.r.t. $x$.
All the three coefficients of the quadratic polynomial ${\bf P^i}_{m,j,k}(x)$ are obtained uniquely by
imposing the conservation of the three volume averages,
$\overline {\bf U}_{m-1,j,k}$, $\overline {\bf U}_{m,j,k}$ and $\overline {\bf U}_{m+1,j,k}$, where $m \in \{i-1, i, i+1\}$.
Thus, each polynomial,
${\bf P^i}_{m,j,k}(x)$, can be written as,
\begin{eqnarray} \label{P_m}
\nonumber {\bf P^i}_{m,j,k}(x) &=& \overline {\bf U}_{m,j,k} \\ \nonumber &&- \frac{1}{24}(\overline {\bf U}_{m+1,j,k}-2\overline {\bf U}_{m,j,k}+\overline {\bf U}_{m-1,j,k}) 
 \\ \nonumber &&+ \frac{\overline {\bf U}_{m+1,j,k} -\overline {\bf U}_{m-1,j,k}}{2 \Delta x}(x-x_m)  
 \\ \nonumber  &&+ \frac{(\overline {\bf U}_{m+1,j,k}-2\overline {\bf U}_{m,j,k}+\overline {\bf U}_{m-1,j,k})} {2\Delta x^2} (x-x_m)^2, 
\\ &{\mathrm{where}}&  m \in \{i-1, i, i+1\}.
\end{eqnarray}
After we reconstruct all the polynomials (${\bf P^i}_{i-1,j,k}$, ${\bf P^i}_{i,j,k}$, ${\bf P^i}_{i+1,j,k}$), smoothness indicators
can easily be computed using Eq.\eqref{ISn}. Nevertheless, we provide here the final expressions of the same for reference.
 \begin{eqnarray}\label{ISn_cweno4}
 \nonumber IS^i_{i-1,j,k} &=& \frac{13}{12} (\overline {\bf U}_{i-2,j,k} - 2 \overline {\bf U}_{i-1,j,k} 
+ \overline {\bf U}_{i,j,k})^2 \\ && + \frac{1}{4} (\overline {\bf U}_{i-2,j,k}-4\overline {\bf U}_{i-1,j,k}+3\overline {\bf U}_{i,j,k})^2, \\ 
 \nonumber IS^i_{i,j,k} &=& \frac{13}{12} (\overline {\bf U}_{i-1,j,k} - 2 \overline {\bf U}_{i,j,k} 
+ \overline {\bf U}_{i+1,j,k})^2 \\ && + \frac{1}{4} (\overline {\bf U}_{i-1,j,k}-\overline {\bf U}_{i+1,j,k})^2, \\ 
 \nonumber IS^i_{i+1,j,k} &=& \frac{13}{12} (\overline {\bf U}_{i,j,k} - 2 \overline {\bf U}_{i+1,j,k} 
+ \overline {\bf U}_{i+2,j,k})^2 \\ && + \frac{1}{4} (3\overline {\bf U}_{i,j,k}-4\overline {\bf U}_{i+1,j,k}+\overline {\bf U}_{i+2,j,k})^2,  
 \end{eqnarray}
Employing the values of the smoothness indicators in Eqs. \eqref{w_l}-\eqref{alpha}, nonlinear weights can be computed.
These weights are finally used to reconstruct the polynomial ${\bf R}_{i,j,k}(x)$ in Eq.\eqref{R_I_4th}.
\begin{figure*}
\includegraphics[width=1.5\columnwidth,center]{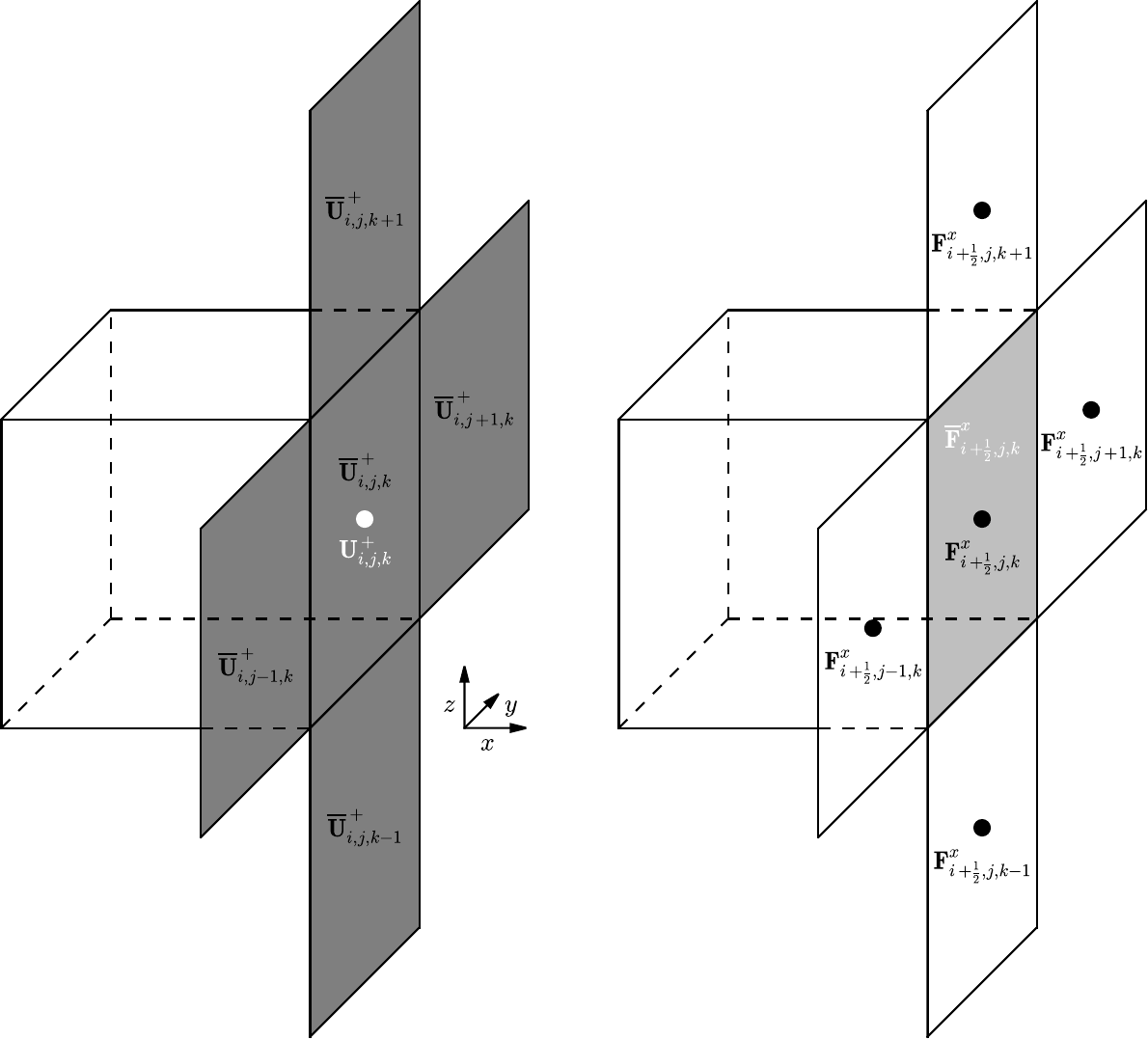}
\caption{ \label{fig:collela_andinv} Illustration of the transformation area-average $\rightarrow$ point value (left) and the inverse transformation point $\rightarrow$ area-average values (right).}
\end{figure*}
\subsection{Step 2: Reconstruction of Area Averages from Volume Averages}
\label{sec:voltoarea}
By construction, the polynomial ${\bf R}_{i,j,k}(x)$ satisfies the conservation constraint,
 \begin{equation}\label{R_ijk}
 \overline {\bf U}_{i,j,k} = \frac{1}{\Delta x} \int_{x_{i-1/2}}^{x_{i+1/2}}\mathrm{d}x  \,  {{\bf R}_{i,j,k}(x)}\ .   
 \end{equation}
Now comparing Eq.\eqref{u_avg} and Eq.\eqref{R_ijk} we obtain,
 \begin{eqnarray}\label{R2_ijk}
\nonumber {{\bf R}_{i,j,k}(x_{i\pm1/2})} &=& \frac{1}{\Delta y\Delta z} \int_{z_{k-1/2}}^{z_{k+1/2}} \int_{y_{j-1/2}}^{y_{j+1/2}} \mathrm{d}y \, \mathrm{d}z \, {{\bf U}}(x_{i\pm1/2},y,z) \, \\ &=& \overline{\bf U}_{i,j,k}^\pm (say) .   
 \end{eqnarray}
For the sake of convenience,
we provide here the formula, all derivation done, for $\overline{\bf U}^{\pm}_{i,j,k}$ from the knowledge of the surrounding cell averages:
 \begin{eqnarray}\label{U+-}
\nonumber \overline{\bf U}_{i,j,k}^+ &=& \frac{1}{6} \Big [ w^i_{i-1,j,k}(11\overline {\bf U}_{i,j,k}-7\overline {\bf U}_{i-1,j,k}+2\overline {\bf U}_{i-2,j,k})\\
\nonumber &&+w^i_{i,j,k}(2\overline {\bf U}_{i+1,j,k}+5\overline {\bf U}_{i,j,k}-\overline {\bf U}_{i-1,j,k})\\
&&+w^i_{i+1,j,k}(2\overline {\bf U}_{i,j,k}+5\overline {\bf U}_{i+1,j,k}-\overline {\bf U}_{i+2,j,k}) \Big ], \\ 
 \nonumber \overline{\bf U}_{i,j,k}^- &=&\frac{1}{6}\Big [ w^i_{i-1,j,k}(2\overline {\bf U}_{i,j,k}+5\overline {\bf U}_{i-1,j,k}-\overline {\bf U}_{i-2,j,k})\\
 \nonumber &&+w^i_{i,j,k}(2\overline {\bf U}_{i-1,j,k}+5\overline {\bf U}_{i,j,k}-\overline {\bf U}_{i+1,j,k})\\
\nonumber &&+w^i_{i+1,j,k}(2\overline {\bf U}_{i+2,j,k}-7\overline {\bf U}_{i+1,j,k}+11\overline {\bf U}_{i,j,k}) \Big ].\\
&&
 \end{eqnarray}
From Eq.\eqref{R2_ijk} it is apparent that the reconstruction polynomial ${\bf R}_{i,j,k}$ does not
provide point values but area-averaged values at the grid cell interfaces.
When employing a Taylor expansion to expand the
polynomial ${{\bf U}}(x_{i\pm1/2},y,z)$ about the face centres $(x_{i\pm1/2},y_j,z_k)$ we obtain,
 \begin{eqnarray}\label{R3_ijk}
\nonumber \overline{\bf U}_{i,j,k}^\pm  &=& {{\bf U}}(x_{i\pm1/2},y_j,z_k) 
                                  + \frac{\Delta y^2}{24} 
\partial_{yy}{{\bf U}}(x_{i\pm1/2},y_j,z_k) \\ \nonumber &&+ \frac{\Delta z^2}{24} 
\partial_{zz}{{\bf U}}(x_{i\pm1/2},y_j,z_k) \\ &&+ {\bf\mathcal O}(\Delta x^4 + \Delta y^4 + \Delta z^4).   
 \end{eqnarray}
This implies that the area averages $\overline{\bf U}_{i,j,k}^\pm$ are second-order
approximations to the point values at the face centres, {\it i.e.}
 \begin{eqnarray}\label{R4_ijk}
\overline{\bf U}_{i,j,k}^\pm  = {{\bf U}}(x_{i\pm1/2},y_j,z_k) 
                                   + {\bf\mathcal O}(\Delta x^4 + \Delta y^2 + \Delta z^2) .   
 \end{eqnarray}
Since the averages cannot preserve their accuracy when subject to multiplication and division, we need to obtain higher-order point values of the physical quantities to compute higher-order fluxes and to maintain the higher-order accuracy of the numerical scheme. In the next subsection, we discuss a method to achieve higher-order point values at the face centres.
\vspace{-0.2in}
\subsection{Step 3: From Area Averages to Point Values}
\label{sec:facetopoint}
In our previous work, we have discussed various methods to compute higher-order point values from the area averages,
$\overline{\bf U}_{i,j,k}^\pm$ \citep{verma2017higher}.
In this work, we use the most efficient method presented in that paper, which is method {\it C} and employs the least number of
reconstruction operations. It is based on \cite{ppm}. This method has also been extended to construct a seventh order finite volume WENO scheme \citep{BUH14}. We compute higher-order accurate point values $\bf U^{\pm}_{i,j,k}$ of the conserved quantities at the face centres from the corresponding area averages $\overline{\bf U}_{i,j,k}^\pm$. This is done through the following formula which is illustrated in figure \ref{fig:collela_andinv}:
\beqa
\label{eq:colella}
\nonumber {{\bf U}}_{i,j,k}^\pm &=&   \overline{\bf U}_{i,j,k}^\pm -\frac{1}{24} ( \overline{\bf U}_{i,j-1,k}^\pm
                                - 2 \overline{\bf U}_{i,j,k}^\pm + \overline{\bf U}_{i,j+1,k}^\pm)
                                         \\ \nonumber &&- \frac{1}{24} ( \overline{\bf U}_{i,j,k-1}^\pm
                                - 2 \overline{\bf U}_{i,j,k}^\pm + \overline{\bf U}_{i,j,k+1}^\pm)
                                \\ &&+ {\bf\mathcal O}(\Delta y^4 + \Delta z^4).
\eeqa
We note here that since the mathematical origin of the relation in Eq.\eqref{eq:colella} is a Taylor expansion,
this transformation is strictly speaking only valid in smooth regions. Section \ref{sec:negcheck3} addresses this point.
\subsection{Step 4: Determination of Point-Valued Fluxes}
\label{sec:pointfluxes}
There are several methods to estimate the physical fluxes at the interfaces by employing various Riemann-solvers.
In this work, we make use of a
simple and robust approximation to the Riemann problem, {\it i.e.} the local Lax-Friedrichs flux (LLF), e.g.,  \citep{shu1989efficient,BUH14,ROM15,verma2017higher}.
The LLF approximation to the point-valued flux at the centre of  a cell interface $(x_{i+1/2},y_j,z_k)$ is given by,
 \begin{eqnarray}\label{Fx}
\nonumber {\bf F^x}_{i+1/2,j,k} &=&  \frac{{\bf f^x}({\bf U}_{i+1,j,k}^-,{\bf B}_{i+1,j,k}^-)
+{\bf f^x}({\bf U}_{i,j,k}^+,{\bf B}_{i,j,k}^+)}{2} \\ &&-  
                              \frac{a^x_{i+1/2,j,k}}{2}  ({\bf U}^{-}_{i+1,j,k}-{\bf U}^{+}_{i,j,k}), 
 \end{eqnarray}
where the quantities $({\bf U}_{i,j,k}^+,{\bf B}_{i,j,k}^+)$ and $({\bf U}_{i+1,j,k}^-,{\bf B}_{i+1,j,k}^-)$  in the present context are fourth-order accurate point values of the conserved quantities at the centre of the grid cell interface $(x_{i+1/2},y_j,z_k)$ and {${\bf f^x}({\bf U}_{i+1/2,j,k}^+,{\bf B}_{i+1/2,j,k}^+)$, ${\bf f^x}({\bf U}_{i+1/2,j,k}^-,{\bf B}_{i+1/2,j,k}^-)$} are the respective point-valued fluxes.

The quantity $a^x_{i+1/2,j,k}$ is the local maximum speed of propagation of information in the $\vx$-direction, estimated as:
\beq
a^x_{i+1/2,j,k}=\max\left( (|v_x|+c^x_f)^{+}_{i,j,k},(|v_x|+c^x_f)^{-}_{i+1,j,k}\right)
\eeq
with $c^x_f$ the magneto-sonic speed:
\beq
c^x_f=\sqrt{\frac{1}{2}\left( (c_s^2+c_A^2)+\sqrt{(c_s^2+c_A^2)^2-4c_s^2\frac{B_x^2}{\rho}} \right)},
\eeq
where $c_s=(\gamma p/\rho)^{\frac{1}{2}}$ is the sound speed and $c_A=(|\vb^2|/\rho)^{\frac{1}{2}}$ is the \alfven speed. We note here that the expression for the point-valued LLF at the face centre of the opposite interface $(x_{i-1/2},y_j,z_k)$, ${\bf F^x}_{i-1/2,j,k}$ can be recovered by employing the operation $i\rightarrow i-1$ in Eq.\eqref{Fx}.
\subsection{Step 5: From Point-Valued Fluxes to Area-Averaged Fluxes}
\label{sec:pointtoface}
Once the point-valued fluxes are determined, one can use the ``inverse" of relation \eqref{eq:colella} in order to compute
fourth-order area-averaged fluxes from the point-valued fluxes \citep{ppm,BUH14,verma2017higher}. These fluxes are the ones appearing in equation \eqref{eq:compactU}. In the $\vx$-direction, this is :
\beqa
\label{eq:invcolella}
\nonumber {\bf \overline F^x}_{i\pm1/2,j,k} &=&   {\bf F^x}_{i\pm1/2,j,k}  \\ \nonumber &&+ \frac{1}{24} ( {\bf F^x}_{i\pm1/2,j-1,k}
                               - 2 {\bf F^x}_{i\pm1/2,j,k} + {\bf F^x}_{i\pm1/2,j+1,k})
                                        \\ \nonumber &&+ \frac{1}{24} ( {\bf F^x}_{i\pm1/2,j,k-1}
                                - 2 {\bf F^x}_{i\pm1/2,j,k} + {\bf F^x}_{i\pm1/2,j,k+1}) \\ &&+ {\bf\mathcal O}(\Delta y^4 + \Delta z^4) .
\eeqa
This relation is illustrated in figure \ref{fig:collela_andinv}.
\subsection{Step 6: Repeat for the Other Directions}
The previous five subsections describe how to compute the $\vF$ flux appearing in equation \eqref{eq:compactU}. The computation of the two remaining fluxes $\vG$ and $\vH$ is done in an analogous way. After all these terms and the ones involving the magnetic field are computed (see section \ref{sec:CT}), a Runge-Kutta procedure is performed for the time integration, which is detailed in section \ref{sec:timeINT}.
\section{Maintaining the solenoidality of the magnetic field}
\label{sec:CT}
The solenoidality of the magnetic field ($\vnabla \cdot \vb=0$) is an important constraint for MHD simulations. The discretisation inherent to numerics leads to errors in $\vnabla \cdot \vb$, so-called ``numerical monopoles", which typically grow in time, causing unphysical effects \citep{brackbill1980effect,powell1994approximate}. In order to avoid the appearance of numerical monopoles, we chose a CT approach, which keeps $\vnabla \cdot \vb=0$ up to machine precision.
\subsection{The Constrained-Transport Approach}
The CT approach was first proposed in \cite{EVH88} and is used in \citep{LOZ00,LOZ04,balsara2004second,ziegler2004central,gardiner2005unsplit,gardiner2008unsplit,rossmanith2006unstaggered,
helzel2011unstaggered,li2010fourth}, for example. Contrary to the hydrodynamic variables and the total energy density $(\rho,\rho v_x,\rho v_y,\rho v_z,e)$, defined as volume averages, the three magnetic field components are described as area averages on the face normal to the field component direction in the CT framework. 
\begin{figure}
\includegraphics[width=0.75\columnwidth,center]{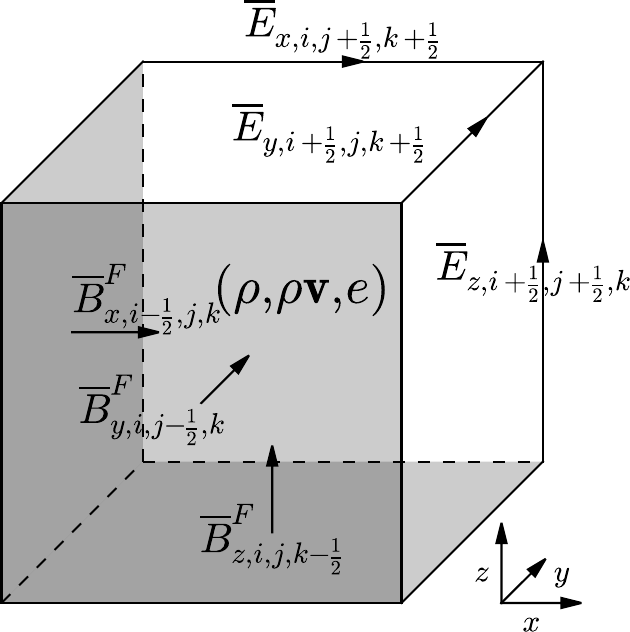}
\caption{ \label{fig:variables} Staggered definition of the magnetic field components, and definition of the edge-averaged electric field component.}
\end{figure}
This staggered definition of the magnetic field is illustrated in figure \ref{fig:variables} and mathematically expressed below:
\beqa
(\overline {B}^F_x)_{i-1/2,j,k}&=&\frac{1}{\Dy\Dz}\int_{\facex}\mathrm{d}y \, \mathrm{d}z \,  B_x(x_{i-1/2},y,z),
\eeqa
\beqa
(\overline B^F_y)_{i,j-1/2,k}&=&\frac{1}{\Dx\Dz}\int_{\facey}\mathrm{d}z \, \mathrm{d}x \,  B_y(x,y_{j-1/2},z),
\eeqa
\beqa
(\overline B^F_z)_{i,j,k-1/2}&=&\frac{1}{\Dx\Dy}\int_{\facez}\mathrm{d}x \, \mathrm{d}y \,  B_z(x,y,z_{k-1/2}).
\eeqa
An application of Stokes' theorem on equation \eqref{eq:mhdb} gives,
\beqa
\label{eq:dtbx}
\nonumber
\frac{d(\overline B^F_x)_{i-1/2,j,k}}{dt}&=&-\frac{(\Ee_z)_{i-1/2,j+1/2,k}-(\Ee_z)_{i-1/2,j-1/2,k}}{\Dy}\\
\nonumber
&&+\frac{(\Ee_y)_{i-1/2,j,k+1/2}-(\Ee_y)_{i-1/2,j,k-1/2}}{\Dz},\\
&&
\eeqa
\beqa
\nonumber
\frac{d(\overline B^F_y)_{i,j-1/2,k}}{dt}&=&\frac{(\Ee_z)_{i+1/2,j-1/2,k}-(\Ee_z)_{i-1/2,j-1/2,k}}{\Dx}\\
\nonumber
&&-\frac{(\Ee_x)_{i,j-1/2,k+1/2}-(\Ee_x)_{i,j-1/2,k-1/2}}{\Dz},\\
&&
\eeqa
\beqa
\label{eq:dtbz}
\nonumber
\frac{d(\overline B^F_z)_{i,j,k-1/2}}{dt}&=&-\frac{(\Ee_y)_{i+1/2,j,k-1/2}-(\Ee_y)_{i-1/2,j,k-1/2}}{\Dx}\\
\nonumber
&&+\frac{(\Ee_x)_{i,j+1/2,k-1/2}-(\Ee_x)_{i,j-1/2,k-1/2}}{\Dy},\\
&&
\eeqa
where the edge-averaged electric field components
play the role of fluxes:

\beqa
(\Ee_x)_{i,j-1/2,k-1/2}=\frac{1}{\Dx}\int_{x_i-\frac{\Dx}{2}}^{x_i+\frac{\Dx}{2}}\mathrm{d}x \, E_x(x,y_{j-1/2},z_{k-1/2}),
\eeqa
\beqa
(\Ee_y)_{i-1/2,j,k-1/2}=\frac{1}{\Dy}\int_{y_j-\frac{\Dy}{2}}^{y_j+\frac{\Dy}{2}}\mathrm{d}y \, E_y(x_{i-1/2},y,z_{k-1/2}),
\eeqa
\beqa
(\Ee_z)_{i-1/2,j-1/2,k}=\frac{1}{\Dz}\int_{z_k-\frac{\Dz}{2}}^{z_k+\frac{\Dz}{2}}\mathrm{d}z \, E_z(x_{i-1/2},y_{j-1/2},z).
\eeqa

It is straightforward to show the conservation of $\vnabla \cdot \vb$ in the second-order approximation, as in \cite{ziegler2004central}. This is because the $\vnabla \cdot \vb$ term in the second-order approximation can be expressed as:
\beqa
\label{eq:2divb}
\nonumber (\vnabla \cdot \vb)_{i,j,k} &\approx& \frac{(\overline B^F_x)_{i+1/2,j,k}-(\overline B^F_x)_{i-1/2,j,k}}{\Dx}\\
\nonumber &&+ \frac{(\overline B^F_y)_{i,j+1/2,k}-(\overline B^F_y)_{i,j-1/2,k}}{\Dy} \\ &&+ \frac{(\overline B^F_z)_{i,j,k+1/2}-(\overline B^F_z)_{i,j,k-1/2}}{\Dz}.
\eeqa
If we take the time derivative of equation \eqref{eq:2divb} and then employ equations \eqref{eq:dtbx}-\eqref{eq:dtbz},
$\frac{d}{dt}(\vnabla \cdot \vb)$ vanishes because the R.H.S. terms cancel pairwise. This means that if the divergence of $\vb$ is zero initially, it remains zero, up to machine precision at all times.
We note here that the accuracy of the magnetic field components depends on the following:
\begin{enumerate}
\item Accuracy of the time integrator employed to evolve the equations \eqref{eq:dtbx}-\eqref{eq:dtbz} in time (discussed in section \ref{sec:timeINT}).
\item Accuracy of the edge-averaged electric field components.
\end{enumerate}
In order to maintain the higher-order accuracy of the scheme, it is hence necessary to compute the edge-averaged electric field components with higher-order accuracy. Besides, we note here that in the CT framework, each magnetic field component is evolved in time as an area-averaged quantity on the faces normal to it. However, in order to compute the fluxes (see Eqs. \eqref{eq:fluxF}-\eqref{eq:fluxh}), reconstruction of the magnetic field components on the other faces is needed as well.
In order to do this, we first compute volume averages of the magnetic field components from the corresponding area averages which are known as primary data. We then treat these volume averages in the same way as the hydrodynamic variables
to perform reconstructions on grid-cell interfaces. Thus, the procedure for computing higher-order edge-averaged electric field components can be described through the following steps:
\begin{enumerate}
\item Construction of volume-averaged magnetic field components from the respective area averages.
\item Deduction of area-averaged magnetic field components on all faces.
\item Computation of area-averaged electric-fluxes at the grid cell interfaces.
\item Reconstruction of edge-averaged electric field components from area-averaged electric-fluxes.
\end{enumerate}
These steps are described in the next subsections.
\subsection{Step 1: Computing Volume Averages for the Magnetic Field Components}
\label{sec:Bcelltoface}
\begin{figure}
\includegraphics[width=\columnwidth,center]{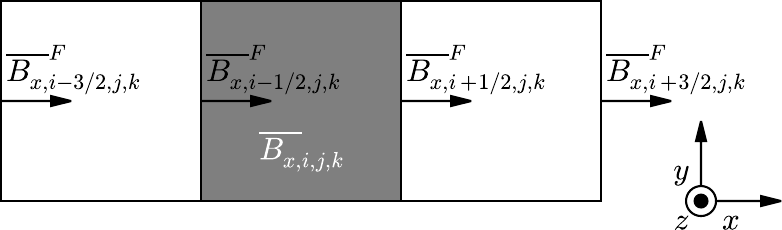}
\caption{ \label{fig:Binterpolation} Illustration of how the volume-averaged magnetic field component along the $\vx$-direction,
$(\overline B_x)_{i,j,k}$ is obtained from the polynomial
interpolation of the neighbouring known area averages, $(\overline B_x^F)$.}
\end{figure}
Although for the determination of the area-average of a variable from the respective volume-average, a fourth-order 1D-CWENO4 reconstruction is performed (see subsections \ref{sec:reconstruction}-\ref{sec:voltoarea}), the area-averaged magnetic field components are continuous along their respective direction (see for e.g. \cite{LOZ00}). Hence, a simple polynomial interpolation is sufficient: one does not need to use a non-oscillatory reconstruction to compute the volume averages from respective area averages. Namely, and as an example, for the grid-cell indexed by $(i,j,k)$ and surrounded by the four faces
$(i\pm3/2,j,k)$ and $(i\pm1/2,j,k)$ where the $\overline B_x^F$ component is known (see figure \ref{fig:Binterpolation}), there is a unique polynomial $Q$ of degree at most three which verifies:
\beqa
\nonumber Q(x-\frac{3\Dx}{2})&=&(\overline B_x^F)_{i-3/2,j,k},
\eeqa
\beqa
\nonumber Q(x-\frac{\Dx}{2})&=&(\overline B_x^F)_{i-1/2,j,k},
\eeqa
\beqa
\nonumber Q(x+\frac{\Dx}{2})&=&(\overline B_x^F)_{i+1/2,j,k},
\eeqa
\beqa
\nonumber Q(x+\frac{3\Dx}{2})&=&(\overline B_x^F)_{i+3/2,j,k}.
\eeqa
Then, a fourth-order approximation of the volume-averaged magnetic field along the $\vx$-direction is deduced from: 
\beq
(\overline B_x)_{i,j,k}=\frac{1}{\Dx}\int_{x_i-\Dx/2}^{x_i+\Dx/2}\mathrm{d}x \, Q(x)
\eeq
In practice, with all the algebra done, this gives:
\beqa
\nonumber (\overline B_x)_{i,j,k}&=&\frac{1}{6}\Big [(\overline B_x^F)_{i-1/2,j,k}+(\overline B_x^F)_{i+1/2,j,k}  \\ &&+\nonumber
\frac{1}{4}(-(\overline B_x^F)_{i-3/2,j,k}+9(\overline B_x^F)_{i-1/2,j,k} \\&&+ 9(\overline B_x^F)_{i+1/2,j,k}-
(\overline B_x^F)_{i+3/2,j,k})\Big].
\eeqa
The computation of the volume-averaged magnetic field components along the $\vy$ and $\vz$-directions is carried out analogously.
\subsection{Step 2: Deducing Area-Averaged Magnetic Field Components on All Faces}
\label{sec:1D-CWENO4bfield}
When the volume-averaged magnetic field components are known, the 1D-CWENO4 procedure described in subsections
\ref{sec:reconstruction}-\ref{sec:voltoarea} is applied and leads ultimately to the knowledge of all the variables on each face in the reconstruction direction [$\pm$]. Note that this procedure is not applied for $B_n$ for a reconstruction in the $n$-direction ($n \in \{x,y,z\}$) since $\overline{B}_n^F$ is already known as primary data.
\subsection{Step 3: Deducing The Area-Averaged Electric Flux}
\label{sec:favE}
In order to understand the origin of electric-fluxes and their role in computing the edge-averaged electric field components we consider, for example, the $x$-component of the equation \eqref{eq:mhdb}:
\beqa
\label{eq:bx}
\pat B_x &=& - \partial_y E_z - \partial_z (-E_y).
\eeqa
Integrating the equation \eqref{eq:bx} over $\Omega_{i,j,k}$ and
applying Gauss' theorem we get:
\beqa
\label{eq:bx-semi}
\nonumber \frac{d (\overline B_x)_{i,j,k}}{dt} &=&  -\frac{\overline{(\Phi_z^{y})}_{i,j+1/2,k}-
\overline{(\Phi_z^{y})}_{i,j-1/2,k}}{\Delta y} \\&&-\frac{\overline{(\Phi_y^{z})}_{i,j,k+1/2}-\overline{(\Phi_y^{z})}_{i,j,k-1/2}} {\Delta z},
\eeqa
here $(\overline B_x)_{i,j,k}$ is the $x$-component of the volume-averaged magnetic field (and not the area-averaged $\overline B_x^F$) and $\overline{(\Phi_y^{z})}_{i,j,k\pm1/2}$
and $\overline{(\Phi_z^{y})}_{i,j\pm1/2,k}$ are the $y$ and $z$-components of the area-averaged electric-fluxes along the $\vz$ and $\vy$-directions, respectively. Analogously, one may write down the semi-discrete equations for $\overline{B_y}$ and $\overline{B_z}$.
Note here that the Eq.\eqref{eq:bx-semi} resembles Eq.\eqref{eq:compactU} and the only difference is that the time evolution of $\overline B_x$ is independent of the (electric) fluxes along the $\vx$-direction. The electric-fluxes are described by the following expressions:
\raggedbottom
\beqa
\label{eq:elecfluxEyz}
\overline{(\Phi_y^{z})}_{i,j,k\pm1/2}&=&\frac{1}{\Dx\Dy}\int_{\facez}  {\mathrm{d}x \, \mathrm{d}y \,[-E_y(x,y,z_{k\pm1/2})]},\\
\label{eq:fluxEzy}
\overline{(\Phi_z^{y})}_{i,j\pm1/2,k}&=&\frac{1}{\Dx\Dz}\int_{\facey} {\mathrm{d}x  \, \mathrm{d}z \, E_z(x,y_{j\pm1/2},z)}.
\eeqa
The remaining expressions for the electric-fluxes can be obtained analogously by considering the $y$ and $z$-components of the
equation \eqref{eq:mhdb} which are:
\beqa
\label{eq:elecfluxEzx}
\overline{(\Phi_z^{x})}_{i\pm1/2,j,k}&=&\frac{1}{\Dy\Dz}\int_{\facex}  {\mathrm{d}y \, \mathrm{d}z \,[-E_z(x_{i\pm1/2},y,z)]},\\
\label{eq:fluxExz}
\overline{(\Phi_x^{z})}_{i,j,k\pm1/2}&=&\frac{1}{\Dx\Dy}\int_{\facez} {\mathrm{d}x  \, \mathrm{d}y \, E_x(x,y,z_{k\pm1/2})},
\eeqa
\beqa
\label{eq:elecfluxEzx}
\overline{(\Phi_x^{y})}_{i,j\pm1/2,k}&=&\frac{1}{\Dz\Dx}\int_{\facey}  {\mathrm{d}z \, \mathrm{d}x \,[-E_x(x,y_{j\pm1/2},z)]},\\
\label{eq:fluxExz}
\overline{(\Phi_y^{x})}_{i\pm1/2,j,k}&=&\frac{1}{\Dy\Dz}\int_{\facex} {\mathrm{d}y  \, \mathrm{d}z \, E_y(x_{i\pm1/2},y,z)}.
\eeqa
Note here that although we compute all the above-mentioned electric-fluxes $\overline{\Phi_n^{m}}$ ($m,n$ stand for $x,y,z$), they are
employed only to compute edge-averaged electric field components in order to evolve the area averages $\overline{B}_n^F, n \in \{x,y,z\}$ in the CT framework. We do not use these fluxes to solve Eq.\eqref{eq:bx-semi} because, as stated above, this method does not conserve the magnetic field solenoidality. In the ideal MHD framework, the electric field is $\vE=-\vv \times \vb$, 
and the point-valued electric field components in the middle of the faces, $E^{\pm}_n$, $n \in \{x,y,z\}$ are deduced from the known point values of $\vv$ and $\vb$. 
In order to compute the area-averaged electric fluxes, when all the variables are known as point values in the middle of each face (see section \ref{sec:facetopoint}), 
point-valued electric-fluxes are first computed using the LLF flux as in section \ref{sec:pointfluxes}. These are given by:
\beqa
\nonumber (\Ef_z^{x})_{i+1/2,j,k}&=&\frac{1}{2}\left({-(E_z)^{-}_{i+1,j,k}-(E_z)^{+}_{i,j,k}}\right)  \\\nonumber&&-\frac{a^x_{i+1/2,j,k}}{2}\left((B_y)^{-}_{i+1,j,k}-(B_y)^{+}_{i,j,k}\right),\\
\eeqa
\beqa
\nonumber (\Ef_y^{x})_{i+1/2,j,k}&=& \frac{1}{2}\left({(E_y)^{-}_{i+1,j,k}+(E_y)^{+}_{i,j,k}}\right)  \\\nonumber&&-\frac{a^x_{i+1/2,j,k}}{2}\left((B_z)^{-}_{i+1,j,k}-(B_z)^{+}_{i,j,k}\right),\\
\eeqa
\beqa
\nonumber (\Ef_z^{y})_{i,j+1/2,k}&=& \frac{1}{2}\left({(E_z)^{-}_{i,j+1,k}+(E_z)^{+}_{i,j,k}}\right)
 \\\nonumber&&-\frac{a^y_{i,j+1/2,k}}{2}\left((B_x)^{-}_{i,j+1,k}-(B_x)^{+}_{i,j,k}\right), \\
\eeqa
\beqa
\nonumber (\Ef_x^{y})_{i,j+1/2,k}&=& \frac{1}{2}\left({-(E_x)^{-}_{i,j+1,k}-(E_x)^{+}_{i,j,k}}\right)  \\\nonumber&&-\frac{a^y_{i,j+1/2,k}}{2}\left((B_z)^{-}_{i,j+1,k}-(B_z)^{+}_{i,j,k}\right), \\
\eeqa
\beqa
\nonumber (\Ef_y^{z})_{i,j,k+1/2}&=& \frac{1}{2}\left({-(E_y)^{-}_{i,j,k+1}-(E_y)^{+}_{i,j,k}}\right)  \\\nonumber&&-\frac{a^z_{i,j,k+1/2}}{2}\left((B_x)^{-}_{i,j,k+1}-(B_x)^{+}_{i,j,k}\right), \\
\eeqa
\beqa
\nonumber (\Ef_x^{z})_{i,j,k+1/2}&=& \frac{1}{2}\left({(E_x)^{-}_{i,j,k+1}+(E_x)^{+}_{i,j,k}}\right)  \\\nonumber&&-\frac{a^z_{i,j,k+1/2}}{2}\left((B_y)^{-}_{i,j,k+1}-(B_y)^{+}_{i,j,k}\right).\\
\eeqa
From these point-valued fluxes $\Phi_n^{m}$, area-averaged electric fluxes $\overline{\Phi_n^{m}}$ are computed using relation \eqref{eq:invcolella}.
\subsection{Step 4: From Area-Averaged Electric Fluxes to Edge-Averaged Electric Field Components}
\label{sec:eavE}
\begin{figure}
\includegraphics[width=0.75\columnwidth,center]{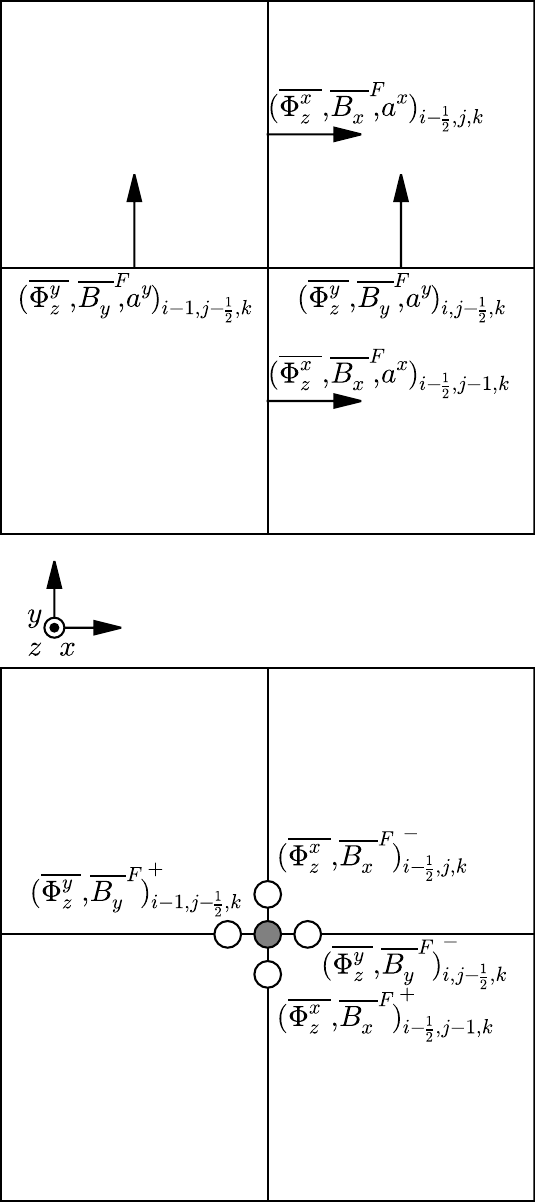}
\caption{ \label{fig:eedge_comp} Illustration of the computation of $\Ee_z$: initially, electric fluxes, magnetic field averages and maximum propagation speeds are known for each faces (top). Then, eight 1D-CWENO4 reconstructions take place, four for the area-averaged electric fluxes giving four relevant edge-averaged values and four for the magnetic field components (bottom). From them, the edge-averaged electric field
component is computed by averaging and taking into account the propagation of discontinuities (filled gray circle).}
\end{figure}
Each edge is surrounded by four faces, on which the area-averaged electric fluxes $\overline{\Phi_n^{m}}$ are known.
Using the same 1D-CWENO4 procedure described in subsections \ref{sec:reconstruction}-\ref{sec:voltoarea}
for each of these four faces leads to four estimates of the edge-averaged electric field components. For the edge-averaged electric field components to plug in the relations \eqref{eq:dtbx}-\eqref{eq:dtbz}, we consider their mean and take into account the propagation of discontinuities. For example, the $z$-component of the edge-averaged electric field, $\Ee_z$ is given by (see figure \ref{fig:eedge_comp}):
\beqa
\nonumber
(\Ee_z)_{i-1/2,j-1/2,k}&\approx& \frac{1}{4}\Big[-(\overline{\Phi_z^{x}})^-_{i-1/2,j,k}-(\overline{\Phi_z^{x}})^+_{i-1/2,j-1,k}
\\ \nonumber &&+ (\overline{\Phi_z^{y}})^-_{i,j-1/2,k}+(\overline{\Phi_z^{y}})^+_{i-1,j-1/2,k}\Big]\\
\nonumber
&&+\frac{\alpha_y}{4}\left({(\overline B_x^F)^-_{i-1/2,j,k}-(\overline B_x^F)^+_{i-1/2,j-1,k}}\right)\\
\label{eq:eez}
\nonumber
&&-\frac{\alpha_x}{4}\left({(\overline{B_y}^F)^-_{i,j-1/2,k}-(\overline{B_y}^F)^+_{i-1,j-1/2,k}}\right),\\
&&
\eeqa
where $(\overline{\Phi_z^{x}})^\pm$ and $(\overline{\Phi_z^{y}})^\pm$ are the edge-averaged electric-fluxes reconstructed from 
the area averages, $(\overline{\Phi_z^{x}})$ and $(\overline{\Phi_z^{y}})$ along
the $\vy$ and $\vx$-directions, respectively and the quantities $(\overline{B}_n^F)^\pm$ are the the edge-averaged magnetic field components reconstructed along the $n$-direction
from area averages $\overline{B}_n^F$, where $n\in \{x,y\}$. This formula can be derived by considering the mean of the two underlying one dimensional LLF fluxes along the $\vx-$ and $\vy-$directions.
We note here that the above expression is different from the one suggested by \cite{LOZ04}, which uses the fields at the diagonal corners. We adapted this formula for reasons of efficiency since computing explicitly the corner terms would require more reconstruction operations. The quantities $\alpha_n$ are estimates of the maximum propagation speed in the $n$-direction, estimated as:
\beqa
\alpha_x&=&\max(a^x_{i-1/2,j,k},a^x_{i-1/2,j-1,k}),\\
\alpha_y&=&\max(a^y_{i,j-1/2,k},a^y_{i-1,j-1/2,k}).
\eeqa
In theory, one should consider the maximum speed among the four reconstructed states. This rough estimation is done for the sake of efficiency since these values are already known from the step described in section \ref{sec:pointfluxes}, as mentioned in \cite{LOZ04}. Other components of the edge-averaged electric field ($\Ee_x, \Ee_y$) can be obtained analogously. These edge-averaged electric field components are the ones used in order to update the magnetic field components through relations \eqref{eq:dtbx}-\eqref{eq:dtbz}.
\section{Negative pressure or density issues}
\label{sec:negative}
Higher-order finite volume schemes, in the vicinity of strong shocks, may lead to unphysical quantities like negative density
or negative pressure during the reconstruction procedure. This issue is a difficulty in varying degrees of severeness for many widely applied numerical solvers, but it is particularly pronounced when employing higher-order precision algorithms. The combination of increased stiffness of higher-order reconstruction polynomials and reduced numerical dissipation
appears to be the culprit. Although numerical diffusion entails amplitude errors in the solution that, for example, reduce small-scale accuracy of the numerical solution, it also tends to smooth out unphysical reconstruction variations. Lower-order numerical schemes thus typically deliver lower accuracy but overall more robust solvers when compared to high-order numerics, in particular for strongly supersonic flows.

Negative density or pressure appears as the higher-order reconstruction polynomial fails to maintain non-oscillatory behaviour around strong shocks. This happens, for example, when more than one strong discontinuity is present in a wide stencil.
Such a configuration poses difficulties for the reconstruction polynomial to choose an oscillation-free reconstruction.
In order to minimise oscillations to increase the robustness of the scheme, we make use of ``global smoothness indicators" (GSI) which are the weighted averages of four smoothness indicators computed from the density and the three magnetic field components. A detailed description of the GSI is provided in Sec. \ref{sec:gsi}. If the scheme still fails to maintain the positivity of the density or the pressure, fallback procedures are employed to reduce the stiffness of the reconstruction polynomial locally during this timestep (cf. sections \ref{sec:negcheck}-\ref{sec:apriori}).

In the extreme case, the fallback procedures can lower the accuracy of the scheme locally down to first-order \citep{mignone2007pluto,tchekhovskoy2007wham,beckwith2011second,ROM15}. Note here that the higher-order accuracy of the scheme remains unaffected in smooth regions under fallback operations because they are employed only around strong discontinuities.
\subsection{Global Smoothness Indicators: Robustness Enhancement and Oscillation Reduction}
\label{sec:gsi}
As mentioned above and noted, for example in \cite{balbas2006nonoscillatory}, the use of individual
smoothness indicators for each variable during the reconstruction leads to significant oscillations in shock regions. We have also observed this in our simulations. Hence, in order to reduce the amount of oscillations, global smoothness indicators (GSI) and global weights, common for all the reconstructed variables, are employed. These GSI can be obtained in various ways, see, e.g. \citep{LPR99,balbas2006nonoscillatory}. The choice of these GSI, employed in the present work, is inspired by \cite{balbas2006nonoscillatory}. The procedure for the same is described below:
\begin{enumerate}
\item Compute the smoothness indicators
for the variables $\rho,B_x,B_y,B_z$.
\item Normalise each smoothness indicator with the corresponding norm of the physical quantity, e.g. the smoothness indicator $IS(\overline \rho)$ based on density $\rho$, needs to be normalised with $||\overline \rho|| = \Big(\Delta x \Delta y \Delta z \sum_{\forall i, j, k} |\overline {\rho}_{i,j,k} |^2\Big)^{1/2}$.
\item Consider the average of all the four normalised smoothness indicators which returns the GSI.
\item Use these GSI in relations \ref{w_l}-\ref{alpha} to deduce the global weights.
\end{enumerate}
For the weights which are used to reconstruct the edge averages from the corresponding area averages in the CT framework
(see section \ref{sec:eavE}), the mean of the GSI surrounding the considered face is taken. For example, given a cell $(i,j,k)$ and the GSI for a reconstruction along the $\vx$-direction $(GSI^{i}_{i,j,k})$, the GSI for a reconstruction from face to edge along the $\vx-$direction of the faces located at $(i,j+1/2,k)$ and $(i,j,k+1/2)$ are:
\beqa
GSI^{i}_{i,j+1/2,k}=\frac{1}{2}(GSI^i_{i,j,k}+GSI^i_{i,j+1,k}), \\
GSI^{i}_{i,j,k+1/2}=\frac{1}{2}(GSI^i_{i,j,k}+GSI^i_{i,j,k+1}).
\eeqa
The weights are then deduced through relations \ref{w_l}-\ref{alpha}. We emphasise here that in addition to reducing the variation of the reconstruction polynomial, the use of global weights reduces the computational cost remarkably. For the numerical method presented in this paper, one indeed performs {\it seven} 1D-CWENO4 reconstructions from volume averages
to area averages and {\it four} 1D-CWENO4 reconstructions from area averages to edge averages for each spatial dimension, that is, {\it eleven} sets of smoothness indicators have to be computed for each spatial dimension if individual weights are computed. With the global weights described in this section, only {\it four} sets of smoothness indicators have to be computed for each spatial dimension, and the weights for the area averages to edge averages are deduced through a low cost averaging of already computed GSI. The influence of this choice of global weights on the oscillation reduction is illustrated in the Brio-Wu shock-tube test (see section \ref{sec:briowu}).
\subsection{A-posteriori Fallback Approach}
\label{sec:negcheck}
In the present method, negative density or negative pressure may appear during the following steps:
\begin{enumerate}
\item Reconstruction procedure.
\item Area average to point value transformation.
\end{enumerate}
Therefore, in the a-posteriori fallback approach, positivity checks are performed after the reconstruction procedure and after the area average to point value transformations. A close look at the
flow-chart of the present algorithm in figure \ref{fig:flowchart} (in section \ref{sec:timeINT}) may be helpful to understand at which points these checks occur. They are detailed in the next subsections. 
\subsubsection{First Positivity Check: after Reconstruction}
\label{sec:negcheck2}
After we reconstruct area averages at the interfaces of a grid-cell, a second-order approximation of the pressure on each face is computed by considering the area-averaged values of the reconstructed quantities as point values. In case the pressure or the density on either of the two faces is negative, the area averages are recomputed using a lower-order reconstruction procedure. Therefore, wherever the 1D-CWENO4 reconstruction fails to give a physical value, we make use of a second-order method, namely the Total-Variation-Diminishing (TVD) slope limiter \citep{van1979towards}. The TVD slope limiter has also been used in a well known MHD code, NIRVANA \citep{ziegler2004central}.
In order to explain the method, we consider the reconstruction along the $\vx$-direction, for example, where the area-averaged
values at the grid cell interfaces are given by:
\beqa
 \bu^{\pm}_{i,j,k}=\bu_{i,j,k} \pm \Delta_x\bu_{i,j,k},
\eeqa
with
\beq
\Delta_x\bu_{i,j,k}=\frac{\max\left[(\bu_{i+1,j,k}-\bu_{i,j,k})\cdot(\bu_{i,j,k}-\bu_{i-1,j,k}),0\right]}{\bu_{i+1,j,k}-\bu_{i-1,j,k}},
\eeq
and analogously in the $\vy$- and $\vz$- directions. In case the resulting density or the second-order approximation for the pressure are still negative, no reconstruction is performed
and we simply take the first-order Godunov scheme approximation $\bu^+_{i,j,k}=\bu^-_{i,j,k}=\bu_{i,j,k}$. In practice, for the standard tests presented in this paper, only the Blast wave (section \ref{sec:blastwave}) and the cloud-shock interaction problem (section \ref{sec:cloudshock}) make use of this a-posteriori lower-order reconstruction. The small amount of cells reconstructed at lower-order, allowing a physical solution at all times, is given in these sections.
\subsubsection{Second Positivity Check: after the Area Average to Point Value Transformation}
\label{sec:negcheck3}
As noted in section \ref{sec:facetopoint}, the area average to point value transformation is strictly speaking only valid in smooth regions. It can happen that as a result of this transformation, negative point values for the pressure or the density are obtained in the vicinity of strong discontinuites. In this case, we consider for this specific face the area averages of all the variables as point values, which corresponds to a physical solution that is a second-order approximation of the point values.
\begin{figure}
\hspace*{-2cm}
\begin{tikzpicture}

\node (magcell) [box] {Computation of magnetic field cell averages (\ref{sec:Bcelltoface})};
\node (rom) [box, below=0pt of magcell] {Computation of reconstruction order map (\ref{sec:apriori})};

\node (loopxyz) [loop, below of=rom, yshift=-0.8cm] {For $n \in \{x,y,z\}$};

\node (reco) [box, right of=loopxyz, xshift=\mar, yshift=-1.5cm] {Reconstruction of all variables but $B_n$ (\ref{sec:reconstruction},\ref{sec:voltoarea},\ref{sec:1D-CWENO4bfield},\ref{sec:apriori_apply})};
\node (check1) [box, below=0pt of reco] {First positivity check (\ref{sec:negcheck2})};
\node (ftop) [box, below of=check1] {Area-to-point transformations (\ref{sec:facetopoint},\ref{sec:apriori_apply})};
\node (check2) [box, below=0pt of ftop] {Second positivity check (\ref{sec:negcheck3})};
\node (pflux) [box, below=0pt of check2] {Computation of point-valued fluxes, including $\Phi$ (\ref{sec:pointfluxes},\ref{sec:favE})};
\node (ptof) [box, below=0pt of pflux] {Point-to-area flux transformation (\ref{sec:pointtoface},\ref{sec:favE},\ref{sec:apriori_apply})};
\node (oneterm) [box, below=0pt of ptof] {Deduction of one term of Eq. \eqref{eq:compactU}};

\node (loopE) [loop, below=0pt of loopxyz,yshift=-8cm] {For $n \in \{x,y,z\}$};

\node (recoE) [box, right of=loopE, xshift=\mar, yshift=-1.5cm] {Reconstruction of $\overline{B}_p^F$ and $\overline{\Phi^p_q}$, for $(p,q) \bot n$ (\ref{sec:eavE})};
\node (Eflux) [box, below=0pt of recoE] {Computation of $\overline{E_n}$ (\ref{sec:eavE})};
\node (twoterms) [box, below=0pt of Eflux] {Deduction of two terms of Eq. \eqref{eq:dtbx}-\eqref{eq:dtbz}};

\node (ApplyRK) [boxRK, below of=loopE, xshift=-1.5cm, yshift=-1.5cm] {Apply Runge-Kutta stage (\ref{sec:timeINT})};

\draw [arrow] (rom) -- (loopxyz);
\draw [arrow] (loopxyz) -|  (reco);
\draw [arrow] (oneterm) -| (loopxyz);

\coordinate [below of=loopxyz,xshift=-2.2cm] (hub1) {};
\coordinate [above of=loopE,yshift=0.2cm] (hub2) {};
\coordinate [below of=loopE,xshift=-2.2cm] (hub3) {};
\coordinate [above of=ApplyRK,yshift=0.2cm] (hub4) {};

\draw [arrow] (loopxyz) -| (hub1) |- (hub2) -- (loopE);

\draw [arrow] (loopE) -| (hub3) |- (hub4) -- (ApplyRK);

\draw [arrow] (loopE) -| (recoE);
\draw [arrow] (twoterms) -| (loopE);

\node (loopxyzdo) [stext]  at ($(loopxyz.east)+(\mardox,\mardoy)$) {do};
\node (loopxyznext) [stext]  at ($(loopxyz.south)+(\marnextx,\marnexty)$) {next};
\node (loopxyzdone) [stext]  at ($(loopxyz.west)+(\mardonex,\mardoney)$) {done};

\node (loopEdo) [stext]  at ($(loopE.east)+(\mardox,\mardoy)$) {do};
\node (loopEnext) [stext]  at ($(loopE.south)+(\marnextx,\marnexty)$) {next};
\node (loopEdone) [stext]  at ($(loopE.west)+(\mardonex,\mardoney)$) {done};

\draw[red,thick,dotted] ($(reco.north west)+(-\marboxx,\marboxy)$)  rectangle ($(oneterm.south east)+(\marboxx,-\marboxy)$);

\node (fluxT) [stextr]  at ($0.5*(reco.north west)+0.5*(oneterm.south west)-(0.5,0)$) {\rotatebox{90}{Interfacial flux computation}};

\draw[red,thick,dotted] ($(magcell.north west)+(-\marboxx,\marboxy)$)  rectangle ($(rom.south east)+(\marboxx,-\marboxy)$);

\node (prepT) [stextr]  at ($0.5*(magcell.north west)+0.5*(rom.south west)-(0.5,0)$) {\rotatebox{90}{Preparation}};

\draw[red,thick,dotted] ($(recoE.north west)+(-\marboxx,\marboxy)$)  rectangle ($(twoterms.south east)+(\marboxx,-\marboxy)$);

\node (EfieldT) [stextr]  at ($0.5*(recoE.north west)+0.5*(twoterms.south west)-(0.5,0)$) {\rotatebox{90}{Constrained Transport}};

\end{tikzpicture}
\caption{ \label{fig:flowchart} Summary of the right-hand side computation. The relevant sections are mentioned in brackets.}
\end{figure}
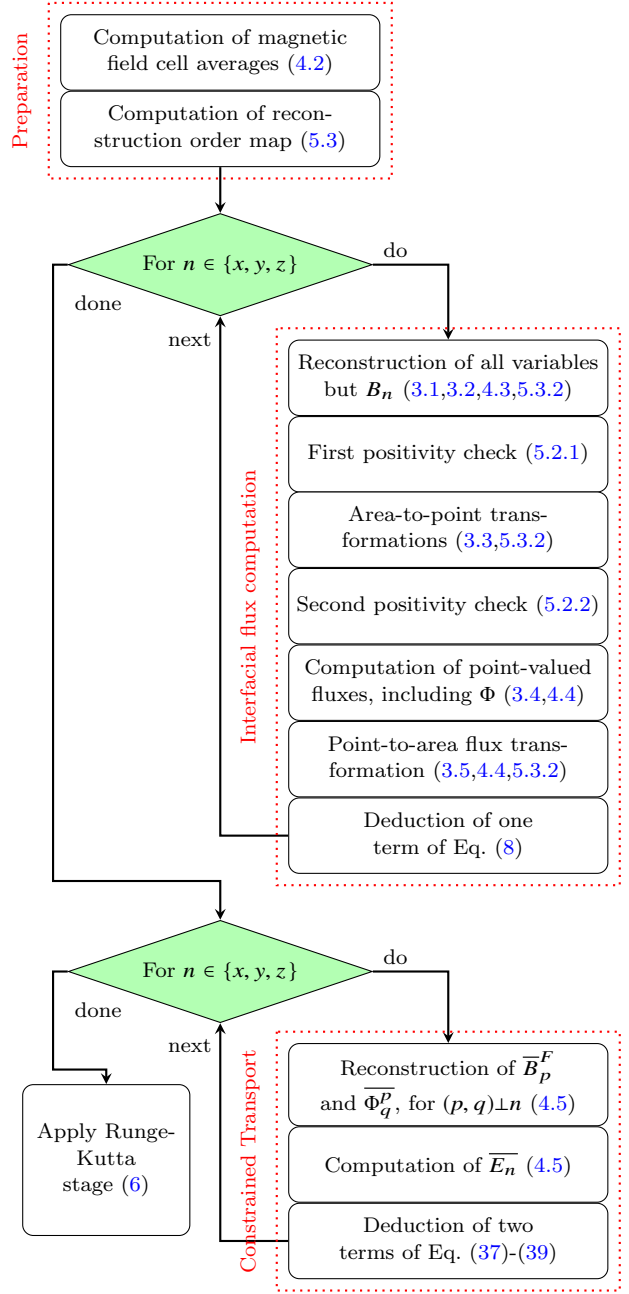

\subsection{A-priori Fallback Approach}
\label{sec:apriori}
In addition to the mechanisms mentioned above, we noticed that for particularly complex problems such as the cloud-shock interaction (see section \ref{sec:cloudshock}), additional dissipation mechanisms are needed in order to ensure that no unphysical solution appears as the system is evolved in time.
In order to achieve this, the reconstructions are computed at lower-order in the vicinity of strong shocks. This approach involves two aspects:
\begin{enumerate}
\item The computation of a ``reconstruction order map" before the main loop for the computation of the right-hand side (see figure \ref{fig:flowchart}).
\item The application of this map for the different steps in the computation of the right-hand side of the MHD equations.
\end{enumerate}
Note that this mechanism is different from the lower-order reconstruction procedure described above since it is applied even if the higher-order method would give a physical state at a certain instant $t$. The implementation of these two aspects is explained in the next subsections.
\subsubsection{Computation of the ``Reconstruction Order Map"}
Additional dissipation should occur only in the vicinity of strong shocks. Several shock indicators exist in the literature. In the present work, we use one inspired by \cite{colella1984piecewise} and based on the pressure gradient, namely: for the cell $(i,j,k)$ and a reconstruction along the $x$-direction, the shock indicator $s^x_{i,j,k}$ is given by:
\beq
        s^x_{i,j,k}=\frac{|p_{i+1,j,k}-p_{i-1,j,k}|}{p_{i,j,k}},
\eeq
with $p_{i,j,k}$ an estimate of the pressure in cell $(i,j,k)$, done by considering the volume averages as if they were point values and plugging them in relation \ref{eq:pressure}. The relation is analogous for the shock indicators along the $\vy$- and $\vz$-directions. Then, for $s^n_{i,j,k}>\eta$, with $\eta$ a certain threshold, the reconstruction order map along the $n$-direction ($n \in \{x,y,z\}$) at cell $(i,j,k)$, $ROM^n_{i,j,k}$, is set to 0 (which corresponds to lower-order reconstruction), otherwise to 1 (high-order reconstruction).

The choice of the threshold $\eta$ depends on the type of the shock indicator and the problem to be solved. For the shock indicator mentioned here, a reasonable ``first guess" for $\eta$ is 4. Indeed, if one would reconstruct naively, without the use of any limiter, the pressure at the cell interfaces by using the second-order approximation $p^{\pm}_{i,j,k}=p_{i,j,k} \pm \frac{p_{i+1,j,k}-p_{i-1,j,k}}{2\Dx}\cdot \frac{\Dx}{2}$, then a value $s^x_{i,j,k}=4$ would already give a pressure equal to zero on one side of the cell. Therefore, we employ a second-order reconstruction method with some limiter, such as TVD, when this threshold is reached. The choice of $\eta$ can also have some impact on the result's symmetry due to the higher order numerics, as described in section \ref{sec:cloudshock}. 
\subsubsection{Local Reduction of the Reconstruction Order}
\label{sec:apriori_apply}
For most of the cells, the reconstruction order map has a value of 1, corresponding to the 1D-CWENO4 reconstruction method. For the cells $(i,j,k)$ where $ROM^n_{i,j,k}=0$, changes occur in the volume-to-area reconstructions (sections \ref{sec:reconstruction}-\ref{sec:voltoarea}), the area averages$\leftrightarrow$point values transformations (sections \ref{sec:facetopoint} and \ref{sec:pointtoface}) and the computation of the edge-averaged electric field components (section \ref{sec:eavE}):
\begin{itemize}
\item For the volume-to-area transformations, the second-order TVD is used for a reconstruction of cell $(i,j,k)$ along the $n$-direction.
\item For area averages$\leftrightarrow$point values transformations, each direction orthogonal to the face normal is considered independently. For example, for a face normal to the $\vx$-direction located at $(i+1/2,j,k)$, the minima of $ROM$ along the $\vy$- and $\vz$-direction for the two cells surrounding the face are respectively considered:
\beqa
\label{eq:ROMyf}
ROM^y_{i+1/2,j,k}&=&\min(ROM^y_{i,j,k},ROM^y_{i+1,j,k})\\
\label{eq:ROMzf}
ROM^z_{i+1/2,j,k}&=&\min(ROM^z_{i,j,k},ROM^z_{i+1,j,k}).
\eeqa
Then the evaluation of the point values ${\bf U}_{i,j,k}^\pm$ from area averages as in relation \ref{eq:colella} is done by:
\beqa
\nonumber {\bf U}_{i,j,k}^\pm & \approx &   \overline{\bf U}_{i,j,k}^\pm -\frac{ROM^y_{i+1/2,j,k}}{24}( \overline{\bf U}_{i,j-1,k}^\pm
                                - 2 \overline{\bf U}_{i,j,k}^\pm + \overline{\bf U}_{i,j+1,k}^\pm)
                                         \\ \nonumber &&- \frac{ROM^z_{i+1/2,j,k}}{24} ( \overline{\bf U}_{i,j,k-1}^\pm
                                - 2 \overline{\bf U}_{i,j,k}^\pm + \overline{\bf U}_{i,j,k+1}^\pm),\\
&&
\eeqa
and similarly for the area average to point value transformation (Eq. \ref{eq:invcolella}). For transformations along the $\vy$- and $\vz$-directions, this is done in an analogous way. This change comes from the fact that the terms involving the second derivatives are only relevant for approximations of order greater than 2 (see Eq. \ref{R4_ijk}).
\item For the computation of the edge-averaged electric field components, two changes occur. First, for the area-to-edge reconstructions, the same procedure as for the volume-to-area reconstructions is used, but using instead of $ROM^n_{i,j,k}$ the minimum of the two $ROM^n_{i,j,k}$ surrounding the considered face (see relations \ref{eq:ROMyf} and \ref{eq:ROMzf}). Second, for the term quantifying the propagation of discontinuities, since second-order schemes do not actually need it, as in \cite{ziegler2004central}, the relation \ref{eq:eez} is changed to:
\beqa
\nonumber
(\Ee_z)_{i-1/2,j-1/2,k}&\approx& \frac{1}{4}\Big[-(\overline{\Phi_z^{x}})^-_{i-1/2,j,k}-(\overline{\Phi_z^{x}})^+_{i-1/2,j-1,k}
\\ \nonumber &&+ (\overline{\Phi_z^{y}})^-_{i,j-1/2,k}+(\overline{\Phi_z^{y}})^+_{i-1,j-1/2,k}\Big]\\
\nonumber
&&+\frac{w^y\alpha_y}{4}\left({(\overline B_x^F)^-_{i-1/2,j,k}-(\overline B_x^F)^+_{i-1/2,j-1,k}}\right)\\
\nonumber
&&-\frac{w^x\alpha_x}{4}\left({(\overline{B_y}^F)^-_{i,j-1/2,k}-(\overline{B_y}^F)^+_{i-1,j-1/2,k}}\right),\\
&&
\eeqa
with $w^x=\min(ROM^x_{i,j-1/2,k},ROM^x_{i-1,j-1/2,k})$ and $w^y=\min(ROM^y_{i-1/2,j,k},ROM^y_{i-1/2,j-1,k})$.
\end{itemize}
As mentioned above, only few cells, faces and edges are affected by these modifications. Section \ref{sec:cloudshock} gives an estimate of their amount.
\section{Time integration and flow-chart}
\label{sec:timeINT}
\begin{table*}
\begin{minipage}{145mm}
\centering
\captionof{table}{Convergence of errors and EOC {for the 2D circularly polarised \alfven wave test, computed from the average of the $L_1$ discrete norms in all the eight variables}. Here spatial averages for all the quantities are compared after one complete period at $t = 0.5$. }
\label{tab:cpaw}
\begin{tabular}{ |cc|c|c|c|c|c|c| }
 \hline
\multicolumn{2}{|c|}{resolution} & $32^2$ & $64^2$ & $128^2$ & $256^2$ & $512^2$ & $1024^2$ \\
\hline
\multirow{2}{*}{$L_1$} & $\delta U_{mean}$ & 1.988 $10^{-4}$ & 7.383 $10^{-6}$ & 3.150 $10^{-7}$ & 1.637 $10^{-8}$ & 9.608 $10^{-10}$& 5.915 $10^{-11}$ \\
                        & EOC      & -               & 4.75             & 4.55 & 4.26 & 4.09 & 4.02 \\
\hline
\end{tabular}
\end{minipage}
\end{table*}

After we compute all the area-averaged fluxes $({\bf \overline F^x},{\bf \overline F^y},{\bf \overline F^z})$ and
edge-averaged electric field components $({\overline E_x},{\overline E_y},{\overline E_z})$, the semi-discrete equations
\eqref{eq:compactU} and \eqref{eq:dtbx}-\eqref{eq:dtbz} are evolved in time using a fourth-order accurate method in order to be consistent with the spatial accuracy.

In order to avoid the introduction of additional oscillations as a part of the time integration process, a Strong Stability-Preserving Runge-Kutta (SSPRK) method is used \citep{shu2009high}. The SSPRK method used in this paper is a ten-stage fourth-order method, described in \cite{SSP10stage}. A pseudocode for a low-storage implementation is available in that paper (pseudocode 3); however, we describe the method here for the sake of completeness. Expressing the R.H.S. of equations \eqref{eq:compactU} and \eqref{eq:dtbx}-\eqref{eq:dtbz} as ${\bf C}[\overline {\bf W}_{i,j,k}]$ where $\overline{\bf W}_{i,j,k} = (\overline {\bf U}_{i,j,k},\overline {\bf B}_{i,j,k}^F)$ and dropping the subscripts $(i,j,k)$, these equations can be rewritten as,
 \begin{equation}\label{semi-dis-new}
 \frac{d\overline {\bf W}(t)}{d t} = {\bf C}[\overline {\bf W}(t)] 
 \end{equation}
The intermediate steps to solve Eq.\eqref{semi-dis-new} are as follows,
 \begin{eqnarray}\label{U1}
   \nonumber   \overline{\bf W}_1 &=&  {{\bf \overline {W}}(t)} + \frac{\Delta t}{6} {\bf C}[{{\bf \overline {W}}(t)}],
 \end{eqnarray}
 \begin{eqnarray}\label{U1}
   \nonumber   \overline{\bf W}_2 &=&  {\bf \overline {W}}_1 + \frac{\Delta t}{6} {\bf C}[{\bf \overline {W}_1}],
 \end{eqnarray}
 \begin{eqnarray}\label{U1}
   \nonumber   \overline{\bf W}_3 &=&  {\bf \overline {W}}_2 + \frac{\Delta t}{6} {\bf C}[{\bf \overline {W}_2}],
 \end{eqnarray}
 \begin{eqnarray}\label{U1}
   \nonumber   \overline{\bf W}_4 &=&  {\bf \overline {W}}_3 + \frac{\Delta t}{6} {\bf C}[{\bf \overline {W}_3}],
 \end{eqnarray}
 \begin{eqnarray}\label{U1}
   \nonumber   \overline{\bf W}_5 &=&  {\bf \overline {W}}_4 + \frac{\Delta t}{6} {\bf C}[{\bf \overline {W}_4}],
 \end{eqnarray}
 \begin{eqnarray}\label{K2}
   \nonumber   {\bf K}_2 &=& \frac{1}{25} {{\bf \overline {W}}(t)} + \frac{9}{25} \overline{\bf W}_5, 
 \end{eqnarray}
 \begin{eqnarray}\label{K2}
   \nonumber   \overline{\bf W}_5 &=& 15 {\bf K}_2 - 5 \overline{\bf W}_5, 
 \end{eqnarray}
 \begin{eqnarray}\label{U1}
   \nonumber   \overline{\bf W}_6 &=&  {\bf \overline {W}}_5 + \frac{\Delta t}{6} {\bf C}[{\bf \overline {W}_5}],
 \end{eqnarray}
 \begin{eqnarray}\label{U1}
   \nonumber   \overline{\bf W}_7 &=&  {\bf \overline {W}}_6 + \frac{\Delta t}{6} {\bf C}[{\bf \overline {W}_6}],
 \end{eqnarray}
 \begin{eqnarray}\label{U1}
   \nonumber   \overline{\bf W}_8 &=&  {\bf \overline {W}}_7 + \frac{\Delta t}{6} {\bf C}[{\bf \overline {W}_7}],
 \end{eqnarray}
 \begin{eqnarray}\label{U1}
   \nonumber   \overline{\bf W}_9 &=&  {\bf \overline {W}}_8 + \frac{\Delta t}{6} {\bf C}[{\bf \overline {W}_8}],
 \end{eqnarray}
 \begin{eqnarray}\label{U1}
   \nonumber   {\overline{\bf W}(t+\Delta t)} &=&  {\bf K}_2 + \frac{3}{5} \overline{\bf W}_9 + \frac{\Delta t}{10} {\bf C}[{\bf \overline {W}_9}].      
\end{eqnarray}
The time step duration $\Dt$ is limited by the maximum propagation speed of any travelling wave through the Courant-Friedrichs-Lewy (CFL) stability criterion:
\beq
\label{eq:cfl}
\Dt = C_{CFL} \underset{i,j,k}{\min}(\Delta x/a^x_{i,j,k},\Delta y/a^y_{i,j,k},\Delta z/a^z_{i,j,k}),
\eeq
where $C_{CFL}$ is the CFL number, see for example \cite{ROM15}.
In the present work, unless stated explicitly, we have chosen a CFL number of $1.95$ for 1D and 2D simulations and $1.55$ for 3D simulations. The numerical method is summarised in the flow-chart shown in figure \ref{fig:flowchart}. 
\begin{table*}
\begin{minipage}{145mm}
\centering
\captionof{table}{Convergence of errors and EOC for the 2D MHD vortex problem, computed from the average of the $L_1$ discrete norms in all the eight variables. Here spatial averages for all the quantities are compared after one complete period at $t = 20$.
}
\label{tab:2dmhdvortex}
\begin{tabular}{ |cc|c|c|c|c|c|c| }
 \hline
\multicolumn{2}{|c|}{resolution} & $32^2$ & $64^2$ & $128^2$ & $256^2$ & $512^2$ & $1024^2$ \\
\hline
\multirow{2}{*}{$L_1$} & $\delta U_{mean}$ & 9.917 $10^{-3}$ & 2.349 $10^{-3}$ & 1.683 $10^{-4}$ & 6.662 $10^{-6}$ & 2.640 $10^{-7}$& 1.315 $10^{-8}$ \\
                        & EOC      & -               & 2.07             & 3.80 & 4.65 & 4.65 & 4.32 \\
\hline
\end{tabular}
\end{minipage}
\end{table*}
\begin{table*}
\begin{minipage}{145mm}
\centering
\captionof{table}{Convergence of errors and EOC for the 3D MHD vortex problem, computed from the average of the $L_1$ discrete norms in all the eight variables. Here spatial averages for all the quantities are compared after one complete period at $t = 10$.
}
\label{tab:3dmhdvortex}
\begin{tabular}{ |cc|c|c|c|c|c|c| }
 \hline
\multicolumn{2}{|c|}{resolution} & $32^3$ & $64^3$ & $128^3$ & $256^3$ & $512^3$\\
\hline
\multirow{2}{*}{$L_1$} & $\delta U_{mean}$ & 4.211 $10^{-4}$ & 4.271 $10^{-5}$ & 1.912 $10^{-6}$ & 7.719 $10^{-8}$ & 4.119 $10^{-9}$ \\
                        & EOC      & -               & 3.30             & 4.48 & 4.63 & 4.22 \\
\hline
\end{tabular}
\end{minipage}
\end{table*}
\begin{table*}
\begin{minipage}{145mm}
\centering
\captionof{table}{Convergence of errors and EOC for the OT vortex problem, computed from the average of the $L_1$ discrete norms in all the eight variables. Here point values for all the quantities are compared at $t = 0.1$ with the reference solution computed at resolution $3232^2$.
}
\label{tab:otvortex}
\begin{tabular}{ |cc|c|c|c|c|c|c| }
 \hline
\multicolumn{2}{|c|}{resolution} & $32^2$ & $96^2$ & $288^2$ & $544^2$ & $800^2$ & $1056^2$ \\
\hline
\multirow{2}{*}{$L_1$} & $\delta U_{mean}$ & 1.569 $10^{-2}$ & 4.000 $10^{-4}$ & 5.087 $10^{-5}$ & 3.862 $10^{-6}$ & 2.552 $10^{-7}$& 1.537 $10^{-8}$ \\
                        & EOC      & -               & 3.340             & 4.03 & 4.38 & 4.27 & 4.23 \\
\hline
\end{tabular}
\end{minipage}
\end{table*}
\section{Numerical tests}
\label{sec:tests} 
\begin{figure*}
\includegraphics[width=2\columnwidth]{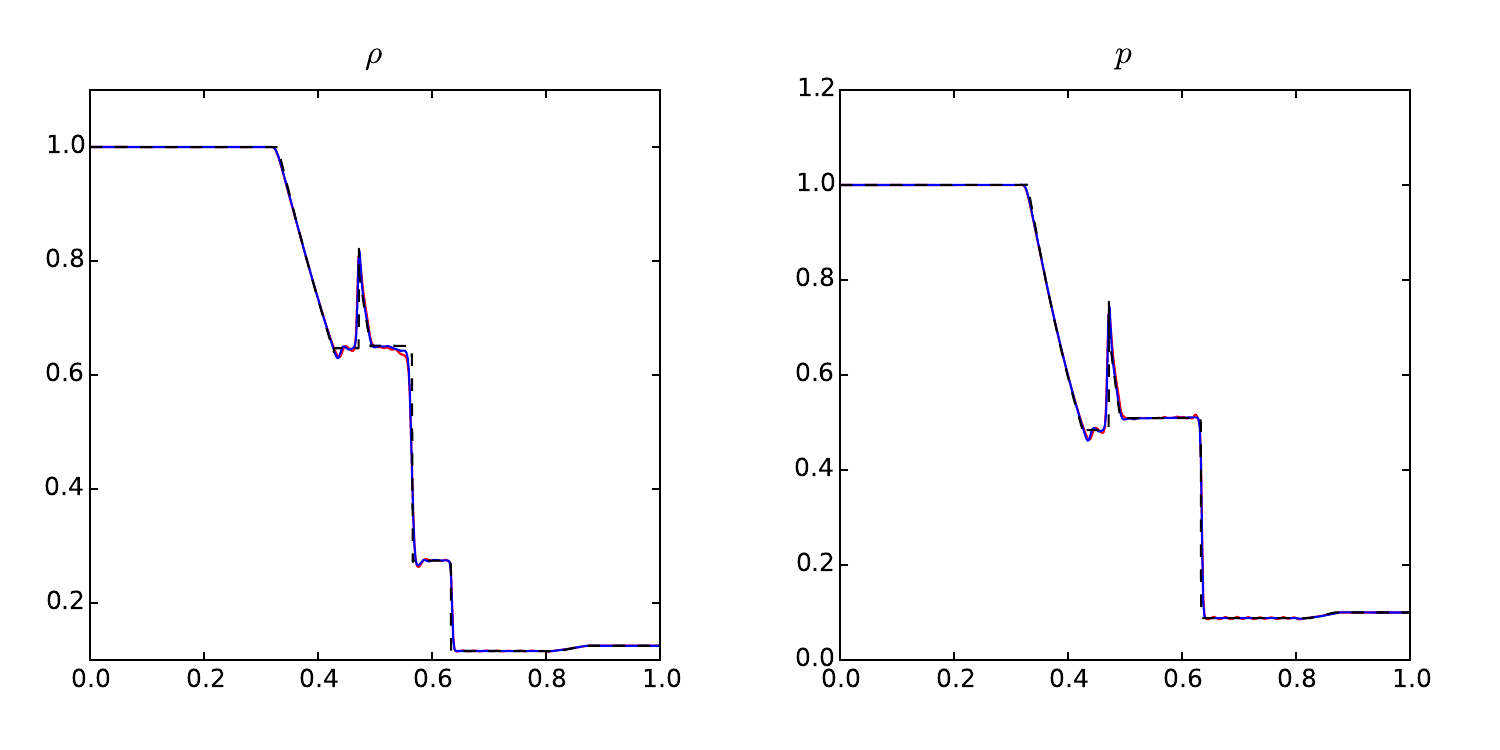}
\caption{ \label{fig:briowu1} 1D Brio-Wu test: the density (left) and the total pressure (right) at time $t = 0.1$ at resolution
$500$ employing global smoothness indicators (solid blue line) and individual smoothness indicators (solid red line). Reference solutions (dashed black line) are obtained at resolution $10000$.}
\end{figure*}
\begin{figure*}
\includegraphics[width=2\columnwidth]{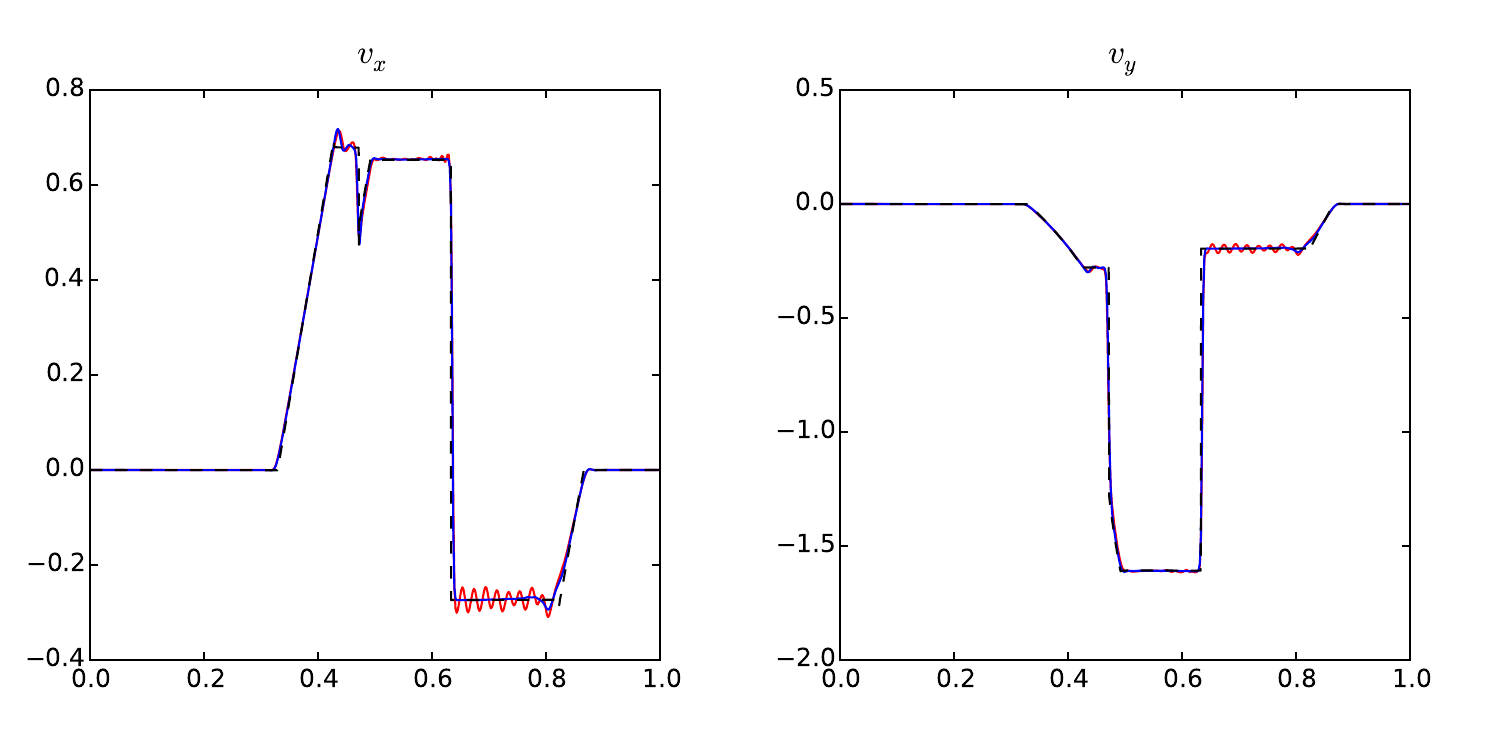}
\caption{ \label{fig:briowu2} 1D Brio-Wu test: the $x$-component of the velocity $v_x$ (left) and the $y$-component of the velocity $v_y$
(right) at time $t = 0.1$ at resolution
$500$ employing global smoothness indicators (solid blue line) and individual smoothness indicators (solid red line). Reference solutions (dashed black line) are obtained at resolution $10000$.}
\end{figure*}
The fourth-order accuracy of the proposed method is confirmed by computing the convergence of errors in various multi-dimensional nonlinear MHD problems, namely: a 2D circularly polarized \alfven wave (CPAW) test \citep{ziegler2004central},
2D and 3D MHD vortex problem \citep{li2010fourth,mignone2010high} and the 2D Orszag-Tang (OT) vortex problem \citep{orszag1979small}. In the CPAW test and MHD vortex problems, convergence studies are performed
after one complete period, however, in the OT vortex problem, it is done at a time when the solution is highly nonlinear but still smooth.

In order to demonstrate the shock capturing behaviour and the robustness of our method we present the results for the 1D Brio-Wu shock-tube test, the evolution of the OT vortex problem, a 2D MHD blast wave \citep{ziegler2004central,ROM15}, the magnetic rotor problem \citep{balsara1999staggered} and the cloud-shock interaction problem \citep{toth2000b,ROM15}.

Lastly, we show some advantages of using higher-order schemes by comparing the effects of numerical dissipation for the second- and fourth-order schemes for a CPAW test, the 2D MHD vortex problem and the blast wave problem in section \ref{sec:dissipation}.

In all the tests, as expected, the solenoidality of the magnetic field remains conserved to machine precision.

\subsection{Accuracy}
In order to estimate the accuracy of the proposed numerical scheme, the error is computed using the $L_1$ discrete norm as
follows \citep{mignone2010high}:
\beq
\label{eq:convERR}
\delta {\bf U}=\frac{\sum_{i,j,k} |{\bf U}^t_{i,j,k}-{\bf U}^{ref}_{i,j,k}|}{N_x N_y N_z},
\eeq
where $\delta {\bf U}$ is the error vector for all the eight MHD variables, ${\bf U}^t$ is the solution at time $t$ and ${\bf U}^{ref}$ is a reference solution. $N_x$, $N_y$ and $N_z$ are the number of grid points along the $\vx$, $\vy$ and $\vz$-directions, respectively.
For 2D tests we keep $N_z = 1$. The mean error is then computed as follows:
\beq
\label{eq:meanerr}
\delta {U_{mean}}=\frac{\sum_{i=1}^8 \delta U(i)}{8},
\eeq
with $i$ the index of one of the eight variables $(\rho,\rho v_x,\rho v_y,\rho v_z,e,B_x,B_y,B_z)$. After computing the errors $\delta {U}_{mean}$ for several resolutions $r_1,r_2,...,r_N$, we obtain the experimental order of convergence $(EOC)$ using the formula,
\begin{equation}
  EOC_i(U) = \frac{|\log\left(\delta U_{mean}(r=r_i)\right)-\log\left(\delta U_{mean}(r=r_{i-1})\right)|}{|\log(r_i)-\log(r_{i-1})|},
\end{equation}
for $i \in \{2,...,N\}$. Since we use fourth-order accurate fluxes and a fourth-order time integration, the expected convergence order is $4$.
\subsubsection{2D Circularly Polarised \alfven Wave Test}
\label{sec:cpaw}
These waves are exact smooth solutions of the MHD equations and therefore they are being used widely to verify the accuracy of a numerical scheme \citep{balsara1999staggered,toth2000b,SGA08,li2010fourth,mignone2010high,ROM15}. In this test the computational domain
is $[0,1]\times[0,1]$, the ratio of specific heats is $\gamma=5/3$ and the initial conditions given by \cite{ziegler2004central}:
\begin{equation}
\begin{pmatrix} \rho \\ v_x \\ v_y \\ v_z \\ p \\ B_x \\ B_y \\ B_z\end{pmatrix} = \begin{pmatrix} 1 \\ -\frac{A}{\sqrt{2}}\sin\left(2\pi(x+y)\right) \\ \frac{A}{\sqrt{2}}\sin\left(2\pi(x+y)\right) \\ A\cos\left(2\pi(x+y)\right) \\ 0.1 \\ \frac{B_0}{\sqrt{2}}+\frac{A}{\sqrt{2}}\sin\left(2\pi(x+y)\right) \\ \frac{B_0}{\sqrt{2}}-\frac{A}{\sqrt{2}}\sin\left(2\pi(x+y)\right) \\ -A\cos\left(2\pi(x+y)\right)\end{pmatrix},
\end{equation}
where we take the wave amplitude $A=0.9$ and the mean-magnetic field along the $x=y$ diagonal $B_0=\sqrt{2}$.
The boundaries are periodic. Under these circumstances, the wave propagates along the diagonal $x=y$ at the \alfven speed associated with the mean-magnetic field: $c_A=\sqrt{\frac{B_0^2}{\rho}}{=\sqrt{2}}$. For the convergence testing, the solution after one complete period {\it i.e.} at time $t=0.5$ is compared to the initial fields, taken as the reference. The test was launched on $32^2, 64^2, 128^2, 256^2, 512^2$ and $1024^2$ grids. The results are shown in table \ref{tab:cpaw} which shows the fourth-order convergence.
\subsubsection{2D MHD Vortex Problem}
\label{sec:mhdvortex}
The initial conditions of this problem, proposed in \cite{balsara2004second}, lead to a magnetised vortex structure propagating along the main diagonal of the computational domain. This structure is a time stationary solution of the nonlinear MHD equations and can be initialised
by the following initial conditions \citep{balsara2004second,ROM15}:
\begin{equation}
\begin{pmatrix} \rho \\ v_x \\ v_y \\ v_z \\ p \\ B_x \\ B_y \\ B_z \end{pmatrix} = \begin{pmatrix} 1 \\ 1-\frac{y\kappa}{2\pi}\exp(\frac{1-r^2}{2}) \\ 1+\frac{x\kappa}{2\pi}\exp(\frac{1-r^2}{2}) \\ 0 \\ 1+\frac{\kappa^2(1-r^2)-\mu^2}{8\pi^2}\exp(1-r^2) \\ \frac{-y\mu}{2\pi}\exp(\frac{1-r^2}{2}) \\ \frac{x\mu}{2\pi}\exp(\frac{1-r^2}{2}) \\ 0 \end{pmatrix},
\end{equation}
with the radius $r=\sqrt{x^2+y^2}$ and $\kappa$ and $\mu$ two parameters we both set to 5 as in \cite{li2010fourth}. The
boundary conditions are periodic. As mentioned in \cite{li2010fourth}, the effects of having a finite domain should be small enough if one wants to observe the proper order of convergence. We hence choose a large domain, $[-10,10]\times[-10,10]$. The ratio of specific heats is chosen to
be $\gamma=5/3$. The vortex propagates at unitary speeds along the $\vx$- and $\vy$-directions, which means that the period of motion is $T=20$. Table \ref{tab:2dmhdvortex} shows the convergence of errors when comparing the solution at $t=T$ with the initial fields and confirms a fourth-order convergence.
\subsubsection{3D MHD Vortex Problem}
\label{sec:3dmhdvortex}
The initial conditions for the 3D MHD vortex were first suggested in \cite{mignone2010high} and are as follows:
\begin{equation}
\begin{pmatrix} \rho \\ v_x \\ v_y \\ v_z \\ p \\ B_x \\ B_y \\ B_z \end{pmatrix} = \begin{pmatrix} 1 \\ 1-{y\kappa}\exp\Big[q(1-r^2)\Big] \\ 1+{x\kappa}\exp\Big[q(1-r^2)\Big] \\ {2} \\ 1+ \frac{1}{4q}\big[\mu^2\big(1-2q(r^2-z^2)\big)-\kappa^2 \rho\big]\exp\Big[2q(1-r^2)\Big] \\ {-y\mu}\exp\Big[q(1-r^2)\Big] \\ {x\mu\exp\Big[q(1-r^2)\Big]} \\ 0 \end{pmatrix},
\end{equation}
with $r$ being the
radius of a sphere, $i.e.$, $r=\sqrt{x^2+y^2+z^2}$.
The parameters are set to $\kappa=\mu=1/(2\pi)$ and $q=1$.
The computational domain is here $[-5,5]\times[-5,5]\times[-5,5]$, $\gamma=5/3$ and $C_{CFL}=1.55$. Table \ref{tab:3dmhdvortex} shows the
convergence of errors by comparing the solution after one period, at $t=10$, with the initial fields. A fourth-order convergence is also observed here.
\begin{figure*}
\includegraphics[width=2\columnwidth]{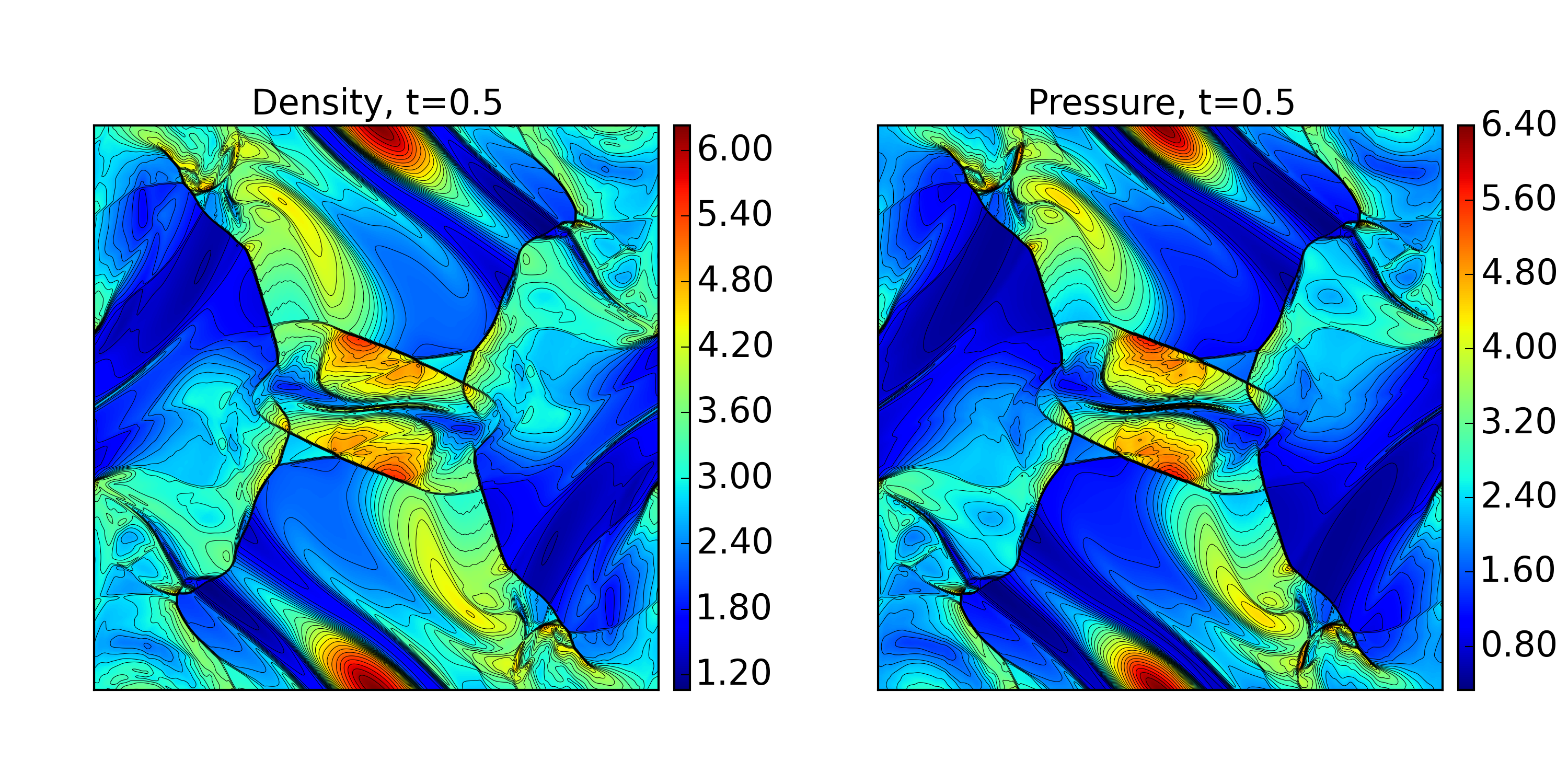}
\caption{ \label{fig:otrob} Orszag-Tang vortex problem: color-coded contour plots of the density (left), the total pressure (right)
at time $t=0.5$ for a resolution $600^2$ are shown.}
\end{figure*}
\begin{figure*}
\includegraphics[width=2\columnwidth]{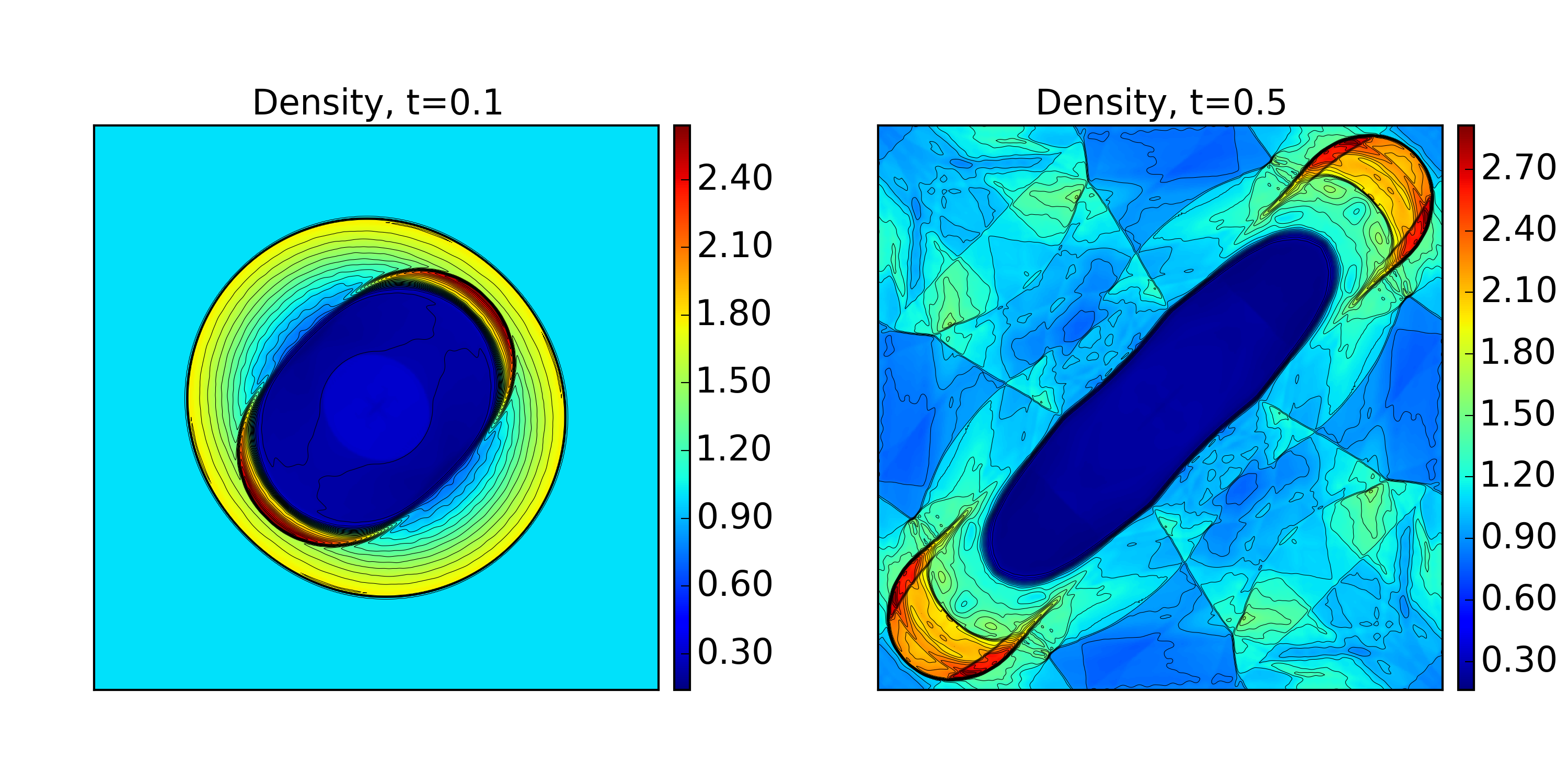}
\caption{ \label{fig:bw1} Blast wave problem: color-coded contour plots of density profiles at times $t=0.1$ (left) and $t=0.5$ (right) at
resolution $600^2$.}
\end{figure*}
\begin{figure*}
\includegraphics[width=2\columnwidth]{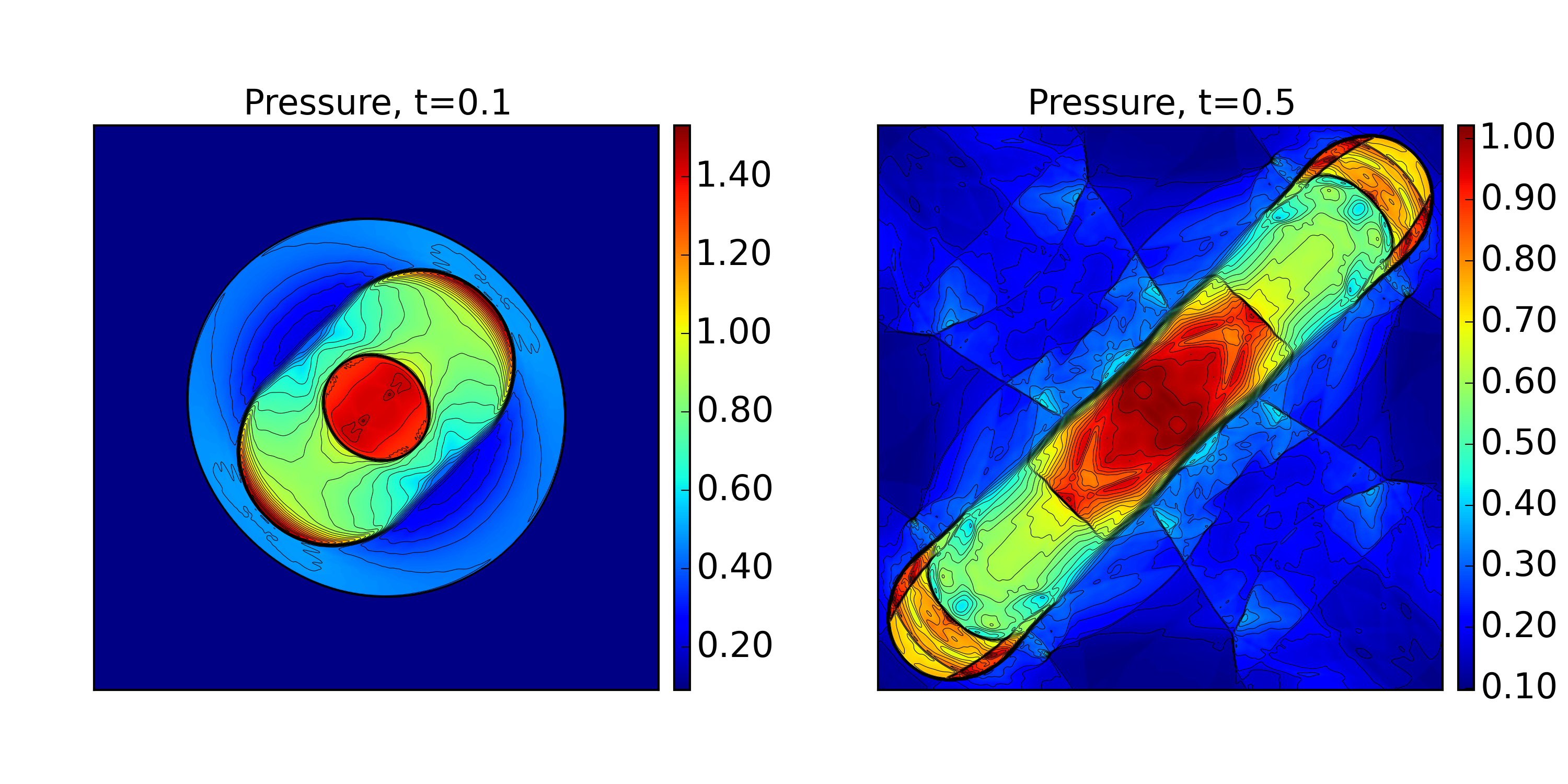}
\caption{ \label{fig:bw2} Blast wave problem: color-coded contour plots of the total pressure profiles at times $t=0.1$ (left) and $t=0.5$ (right) at resolution $600^2$.
}
\end{figure*}
\begin{figure*}
\includegraphics[width=2\columnwidth]{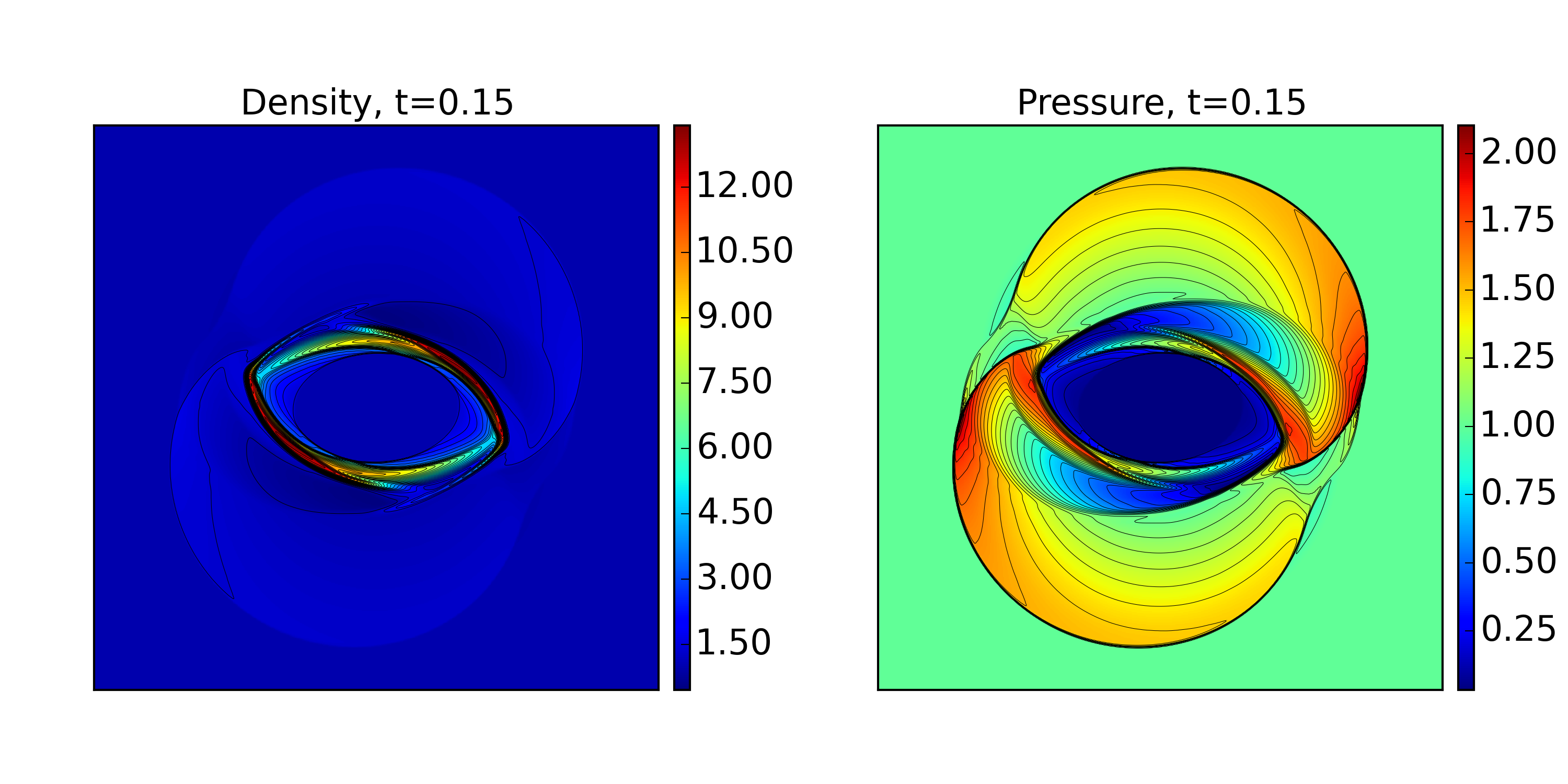}
\caption{ \label{fig:rotor1} The magnetic rotor problem: color-coded contour plots of the density (left), the total pressure (right)
at time $t=0.15$ at resolution $600^2$ are shown.
}
\end{figure*}
\begin{figure*}
\includegraphics[width=2\columnwidth]{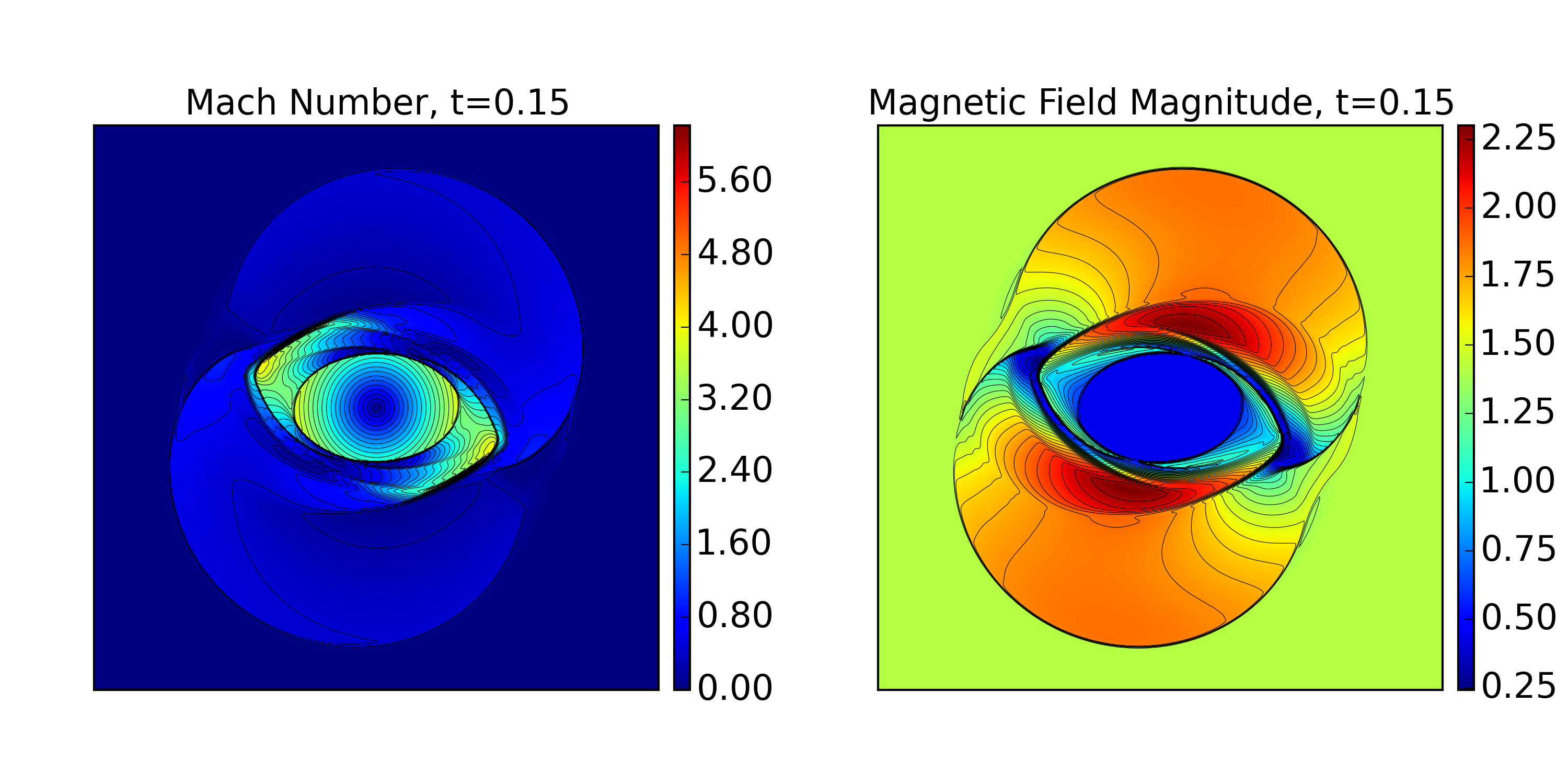}
\caption{ \label{fig:rotor2} The magnetic rotor problem: color-coded contour plots of the Mach number (left), the magnetic field magnitude (right) at time $t=0.15$ at resolution $600^2$ are shown.
}
\end{figure*}
\begin{figure*}
\includegraphics[width=2\columnwidth]{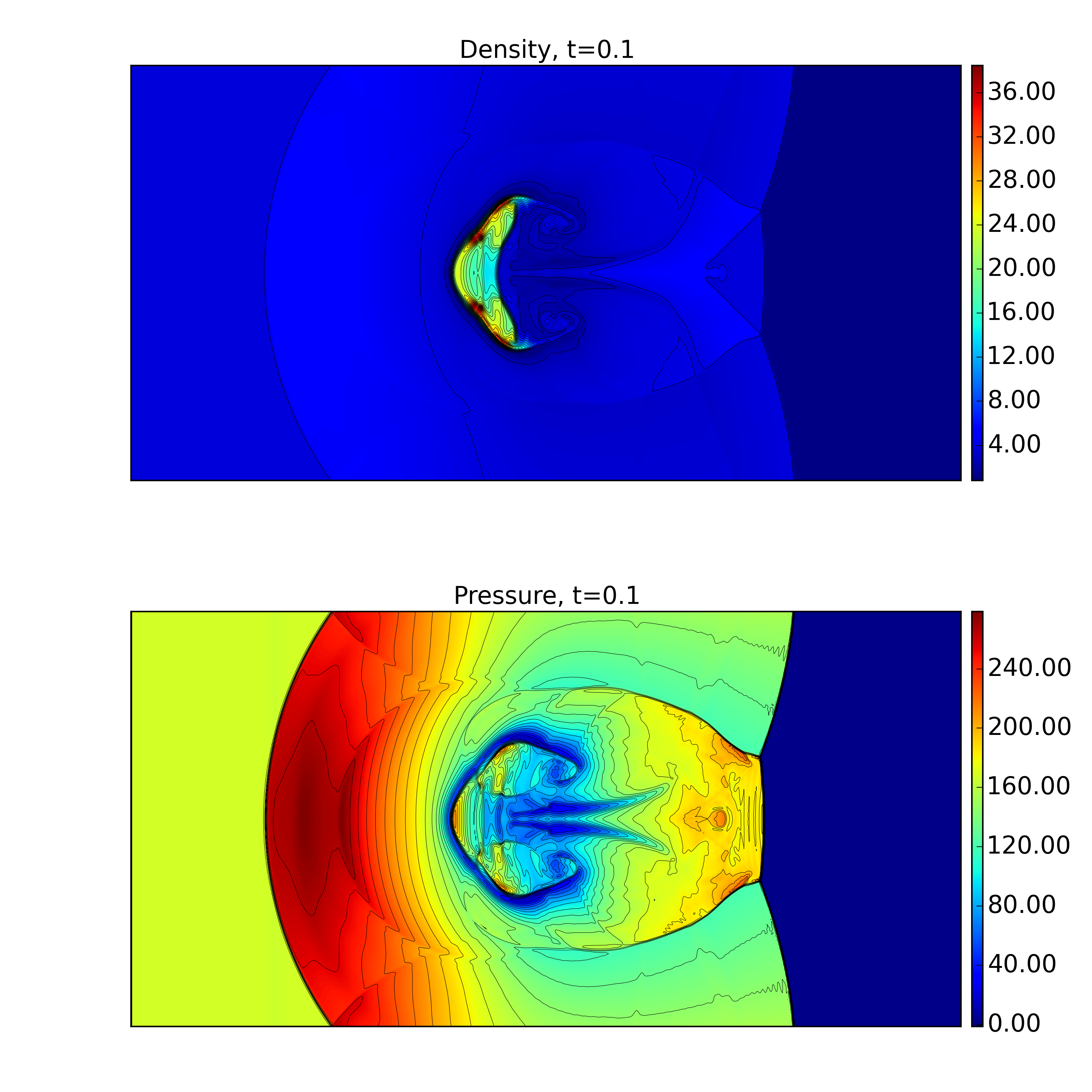}
\caption{ \label{fig:cloudshock} The cloud-shock interaction problem: color-coded contour plots of the density (top),
the total pressure (bottom) at time $t=0.1$ at resolution $800\times400$ are shown.
}
\end{figure*}
\begin{figure*}
\includegraphics[width=2\columnwidth]{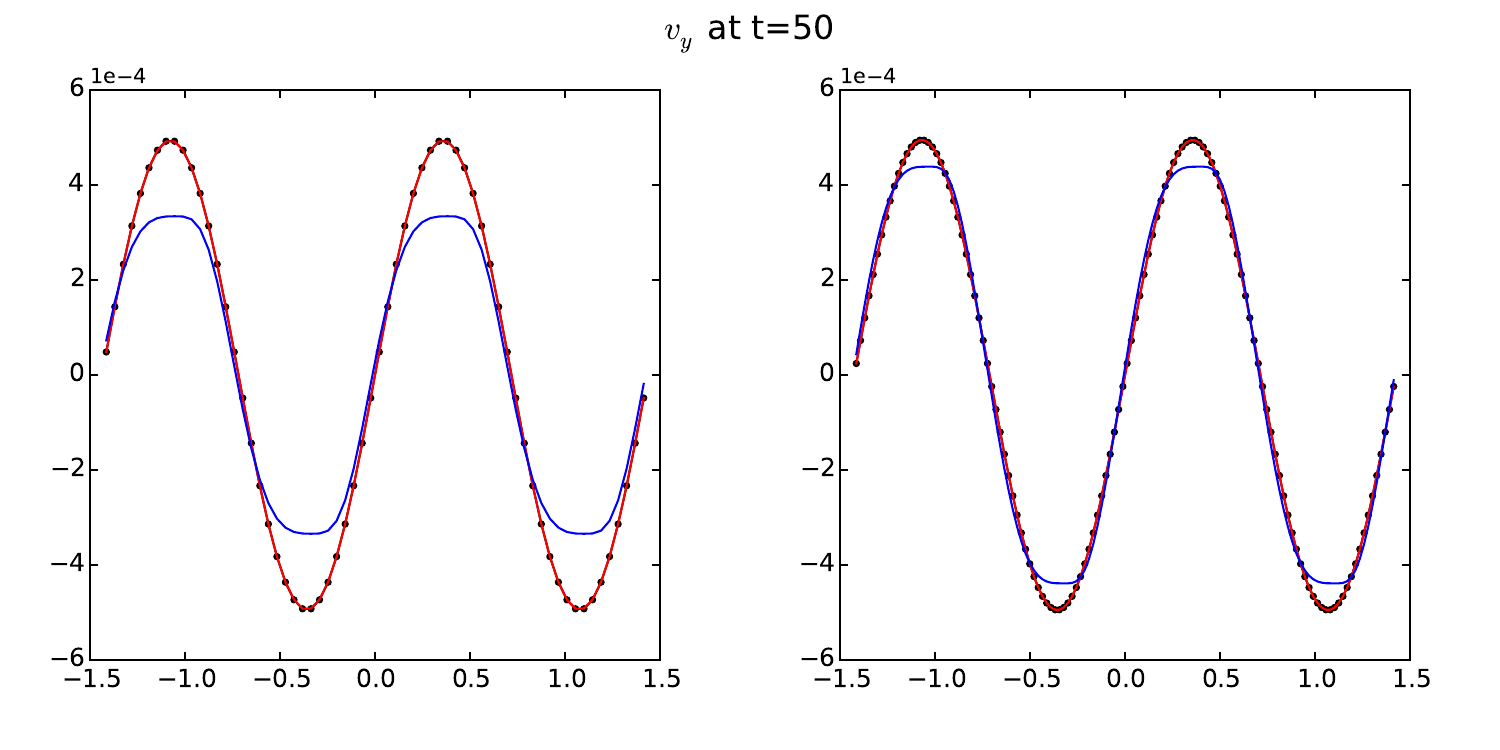}
\caption{ \label{fig:cpawdissi} The 2D CPAW test: 1D plots of the $y$-component of the velocity $v_y$ along
the diagonal at time $t = 50$ ($100$ periods) at resolutions $64^2$ (left) and $128^2$ (right). Here, red lines represent
the results from the fourth-order scheme, blue lines denote the second-order scheme and black line-points
represent the reference solution.
}
\end{figure*}
\begin{figure*}
\includegraphics[width=2\columnwidth]{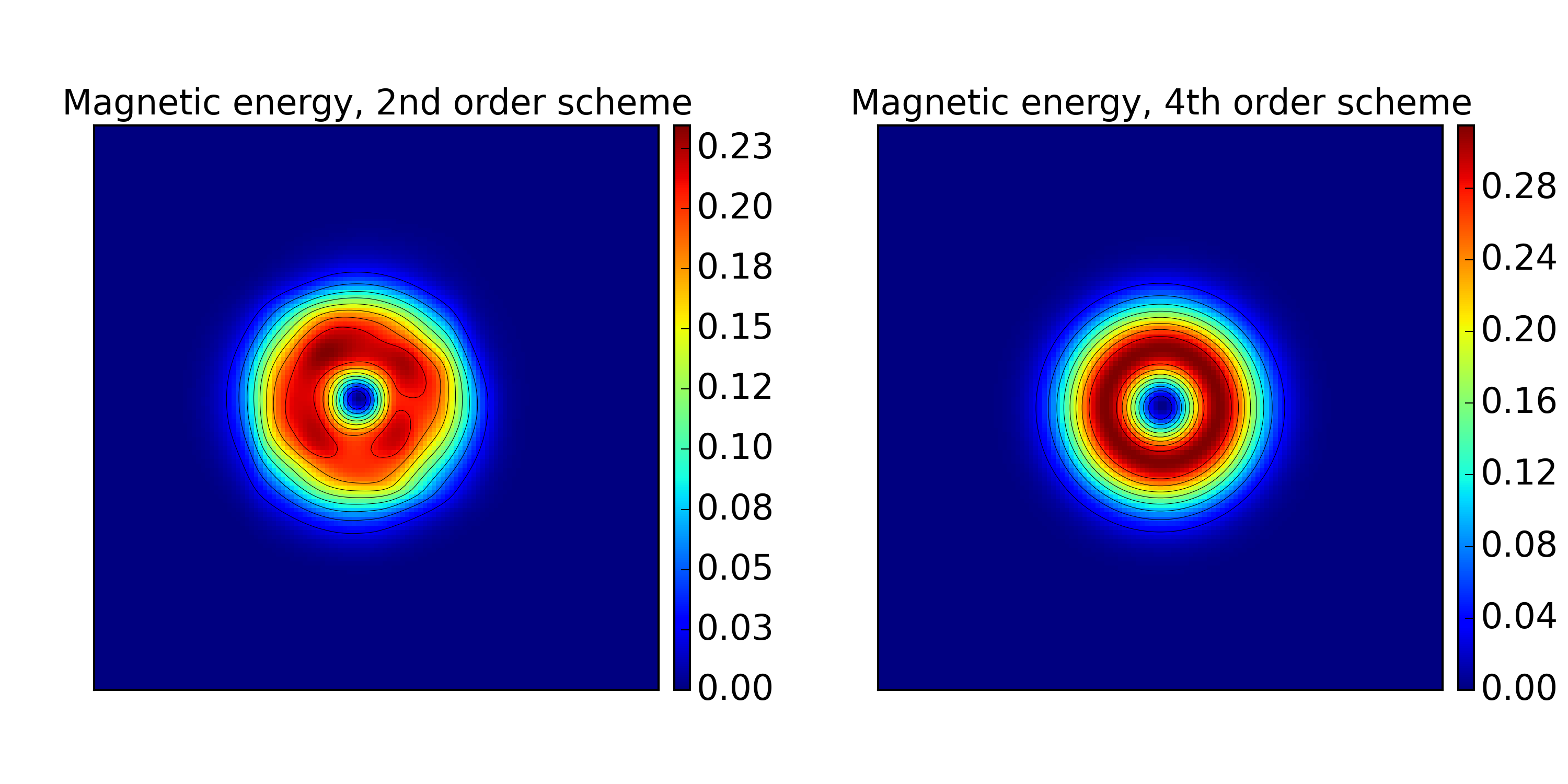}
\caption{ \label{fig:mhdvortexdissi1} 2D MHD vortex problem: color-coded contour plots of the magnetic energy after $10$ periods for a domain size of 10$\times$10 from
the second-order scheme (left) and the
fourth-order scheme (right) at resolution $128^2$.
}
\end{figure*}
\begin{figure}
\includegraphics[width=1\columnwidth]{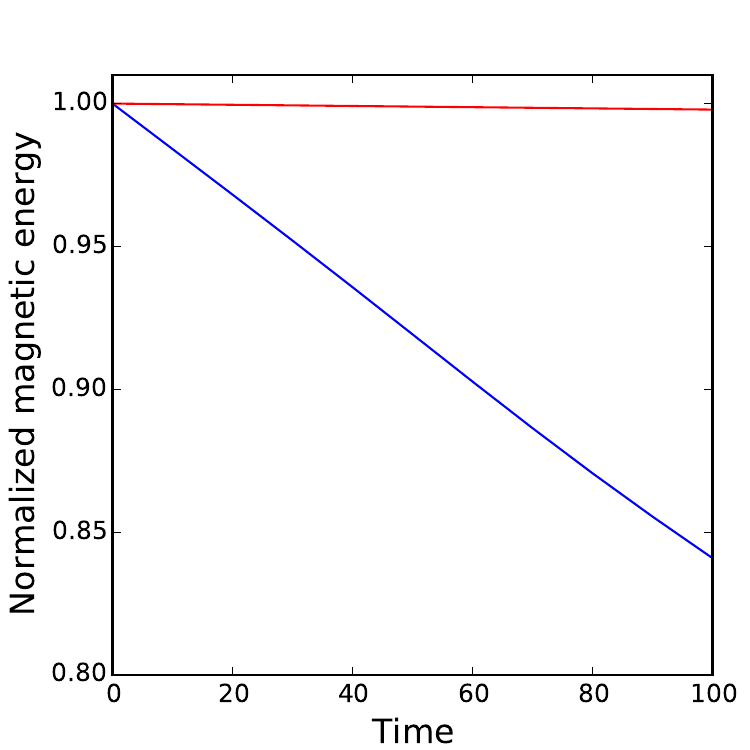}
\caption{\label{fig:mhdvortexdissi2} 2D MHD vortex problem: time evolution of the normalised magnetic energy up to $t = 100$ (ten periods for a domain size of 10$\times$10) at resolution 128$^2$,
where the red-line represents the fourth-order scheme and the blue-line stands for the second-order scheme. The magnetic energy is about 99.8\% of its initial value for the fourth-order scheme.
}
\end{figure}
\begin{figure*}
\includegraphics[width=2\columnwidth]{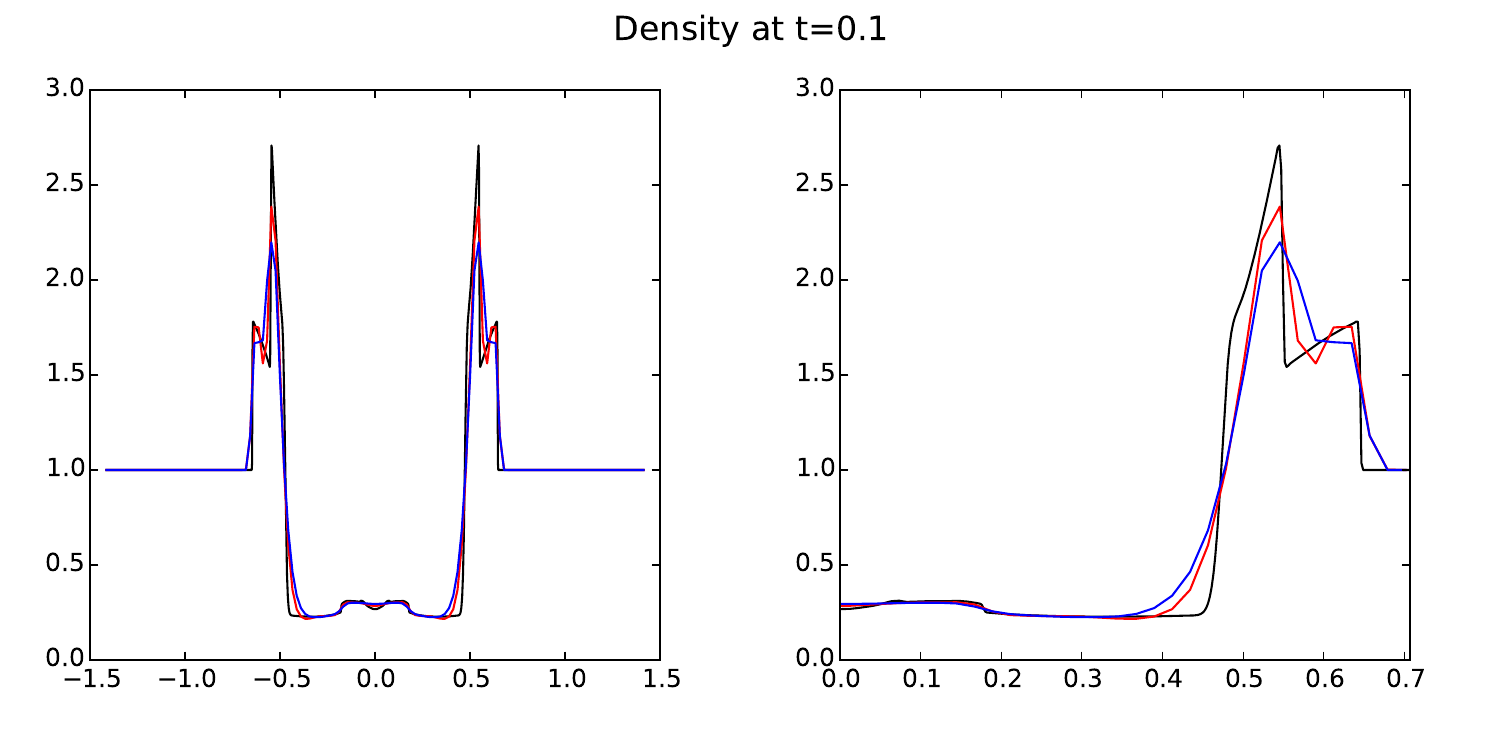}
\caption{ \label{fig:blastwavdissi} MHD blast wave problem: 1D cut of the density profiles along the diagonal of the computational domain at time $t=0.1$ where the left subplot covers the full domain along the diagonal, and the right subplot is zoomed in the range [0,0.7]. The red and blue lines stand for the fourth and second-order schemes, respectively, obtained at resolution $128^2$. The black-line denotes the reference solution obtained from the second-order scheme at
resolution $1280^2$.
}
\end{figure*}
\subsubsection{Orszag-Tang Vortex Problem}
\label{sec:convOT}
The Orszag-Tang (OT) vortex problem, first described in \cite{orszag1979small} in the framework of incompressible MHD equations, has become a standard test to check the robustness of a numerical scheme employed to solve compressible MHD equations. This is because, in this test, the smooth initial conditions lead to the formation of multiple shock structures which end up developing turbulent dynamics.
This problem has been described and solved in several papers, including \citep{toth2000b,ziegler2004central,delzanna:ctdivb,SGA08,li2010fourth,mignone2010high,ROM15}. The initial conditions for this test are given by:
\begin{equation}
\label{eq:initOT}
\begin{pmatrix} \rho \\ v_x \\ v_y \\ v_z \\ p \\ B_x \\ B_y \\ B_z \end{pmatrix} = \begin{pmatrix} \gamma^2 \\ -\sin(2\pi y) \\ \sin(2\pi x) \\ 0 \\ \gamma \\ -\sin(2\pi y) \\ \sin(4\pi x) \\ 0 \end{pmatrix}.
\end{equation}
Periodic boundary conditions are used for the computational domain $[0,1]\times[0,1]$ and 
the ratio of specific heats is $\gamma=5/3$.
As mentioned above, MHD shocks develop quite soon in this test (shown in the next subsection), hence, in order to employ this test for the study of convergence, we choose an instant of time $t=0.1$ where the solution is still smooth. The simulations are launched at resolutions $32^2, 96^2, 288^2, 544^2, 800^2$, $1056^2$ and $3232^2$. The CFL number in this test is chosen to be $1.85$.
For each resolution, $32^2$ point values are extracted, corresponding to the values of the fields at the cell centres
for the lowest resolution, that is $(x=\frac{0.5+i}{32},y=\frac{0.5+j}{32})_{(i,j)\in\{1,32\}}$. These point values are computed by using relation \eqref{eq:colella}, in order to obtain fourth-order point values at the cell centres from the cell averages. The $32^2$ points from the solution launched with resolution $3232^2$
serve as a reference, which are compared to the $32^2$ extracted points from the lower resolutions using relation \eqref{eq:convERR}, but with point values instead of cell averages. Therefore, for this test $N_x = N_y = 32$ and $N_z=1$ remain fixed in Eq. \eqref{eq:convERR} when computing the error at different resolutions. The results are shown in table \ref{tab:otvortex} which demonstrates fourth-order convergence here as well.
\subsection{Shock-Capturing Nature and Robustness}
In order to show the shock capturing nature and robustness of our numerical method, we present five numerical tests: the 1D Brio-Wu shock-tube test, the Orszag-Tang vortex, an MHD Blast Wave, the magnetic rotor problem and the cloud-shock interaction problem. These tests are very demanding and show the ability of the proposed scheme at handling strong and various kinds of MHD shocks.
\subsubsection{1D Brio-Wu Shock-Tube}
\label{sec:briowu}
The Brio-Wu shock-tube problem \citep{brio1988upwind} is an MHD extension of Sod's shock-tube problem in hydrodynamics
\citep{sod1978survey}. This test is usually performed to quantify the ability of a numerical scheme in characterising both continuous and discontinuous flow features.
The initial conditions for this test are as follows:
\begin{equation}
\begin{pmatrix} \rho \\ v_x \\ v_y \\ v_z \\ p \\ B_x \\ B_y \\ B_z \end{pmatrix}_{{x\leq 0.5}} = 
\begin{pmatrix} 1 \\ 0 \\ 0 \\ 0 \\ 1 \\ 0.75 \\ 1 \\ 0 \end{pmatrix}, \qquad 
\begin{pmatrix} \rho \\ v_x \\ v_y \\ v_z \\ p \\ B_x \\ B_y \\ B_z \end{pmatrix}_{{x>0.5}} = \begin{pmatrix} 0.125 \\ 0 \\ 0 \\ 0 \\ 0.1 \\ 0.75 \\ -1 \\ 0 \end{pmatrix}.
\end{equation}
Transmissive boundary conditions are used for the computational domain $[0,1]$ and the ratio of specific heats is $\gamma=5/3$.
In figures \ref{fig:briowu1}-\ref{fig:briowu2}, we show spatial profiles of the density ($\rho$), the total pressure ($p$),
$x$- and $y$-components of the velocities ($v_x$, $v_y$) at the instant of time $t = 0.1$ at resolution $500$. These results are compared with the reference solutions at resolution $10000$.
We note here that these results are comparable to the ones in \cite{ROM15}, however, significant oscillations around the discontinuities can be observed if one employs individual smoothness indicators (compare the blue and red curves in figures \ref{fig:briowu1}-\ref{fig:briowu2}).
\subsubsection{Orszag-Tang Vortex}
The OT vortex problem and the initial conditions are already discussed in section \ref{sec:convOT}. Figure \ref{fig:otrob} shows contour plots of the density and the total pressure at time $t=0.5$.
Here we have chosen to show the total pressure, instead of the magnetic pressure as in \cite{ROM15}, because the latter is contained in the total pressure. We also note here that these solutions are obtained by using only 1D-CWENO4 reconstruction, that is, no cells are reconstructed at lower order to have a physical solution (see section \ref{sec:negative}). This figure is directly comparable to \citep{SGA08,ROM15}, amongst others.
\subsubsection{Blast Wave}
\label{sec:blastwave}
The blast wave problem is solved to show the ability of a numerical scheme at handling the propagation and formation of strong MHD shock waves. Similar tests are described, e.g. in \citep{LOZ00,SGA08,ROM15}. We choose the same initial conditions as in \cite{ROM15}, that is a fluid at rest:
\begin{equation}
\begin{pmatrix} \rho \\ v_x \\ v_y \\ v_z \\ B_x \\ B_y \\ B_z \end{pmatrix} = \begin{pmatrix} 1 \\ 0 \\ 0 \\ 0 \\ 1/\sqrt{2} \\ 1/\sqrt{2} \\ 0 \end{pmatrix},
\end{equation}
for which a large overpressure is present in a cylindrical region: $p=10$ for $r=\sqrt{(x-0.5)^2+(y-0.5)^2}<r_0=0.1$ and $p=0.1$ otherwise.

Periodic boundary conditions are used for the computational domain $[0,1]\times[0,1]$ and the ratio of specific heats is $\gamma=5/3$. Figures \ref{fig:bw1}-\ref{fig:bw2} show color-coded contour plots of the density and the total pressure at the instants $t=0.1, 0.5$ at resolution $600^2$. For this test, even though the a-priori fallback approach is not required to obtain a physical solution at all times, at the very beginning, for $t<0.009$, a minimal amount of cells (of the order of $0.001$--$0.03\%$ of the total number of cells per time-step) is reconstructed at lower-order, and around $0.0001$--$0.01\%$ of the area-to-point transformations are switched off by the a-posteriori fallback approach (see section \ref{sec:negcheck}).
\subsubsection{Magnetic Rotor Problem}
The magnetic rotor problem was suggested for the first time by \cite{balsara1999staggered} which has later been studied by several others, including \citep{toth2000b,ziegler2004central,SGA08,ROM15}. This problem can be initialised as follows: in the centre, for $r=\sqrt{x^2+y^2}<r_0=0.1$, a dense fluid rotates with a constant angular velocity $\omega = 20$. It is embedded in an ambient medium at rest for $r>r_1=0.115$. The ratio of specific heats is $\gamma=7/5$. In order to reduce initial transients, a linear taper is applied for $r$ between $r_0$ and $r_1$, where the density and the velocity norm decrease linearly with $r$:
\begin{eqnarray}
\nonumber \begin{pmatrix} \rho \\ v_x \\ v_y \\ v_z \\ p \end{pmatrix}_{r\leq r_0} = 
\begin{pmatrix} 10 \\ -\omega y \\ \omega x \\ 0 \\ 1 \end{pmatrix},  
\begin{pmatrix} \rho \\ v_x \\ v_y \\ v_z \\ p \end{pmatrix}_{r_0<r\leq r_1} = 
\begin{pmatrix} 1+9f \\ -f\omega yr_0/r \\ f\omega xr_0/r \\ 0 \\ 1  \end{pmatrix},   
\begin{pmatrix} \rho \\ v_x \\ v_y \\ v_z \\ p \end{pmatrix}_{r>r_1} = \begin{pmatrix} 1 \\ 0 \\ 0 \\ 0 \\ 1 \end{pmatrix}, \\&&
\end{eqnarray}
with $f=\frac{r_1-r}{r_1-r_0}$. The magnetic field is everywhere:
\beq
\begin{pmatrix} B_x \\ B_y \\ B_z \end{pmatrix} = \begin{pmatrix} \frac{5}{\sqrt{4\pi}} \\ 0 \\ 0 \end{pmatrix}.     
\eeq
Figure \ref{fig:rotor1} shows contour plots of the density and the total pressure at time $t=0.15$. Figure \ref{fig:rotor2} contains contour plots of the Mach number and the magnetic field magnitude at time $t=0.15$. These solutions are also obtained without the use of fallback approach (see section \ref{sec:negative}). These figures are directly comparable to \cite{SGA08}, amongst others.
\subsubsection{Cloud-Shock Interaction Problem}
\label{sec:cloudshock}
In this problem, we study the interaction between a shock-wave and a high-density cloud. The computational domain is $[0,2]\times[0,1]$ with transmissive boundary conditions and the ratio of specific heats is $\gamma=5/3$. The initial conditions for this test are the same as in \citep{toth2000b,ROM15}:
\begin{equation}
\begin{pmatrix} \rho \\ v_x \\ v_y \\ v_z \\ p \\ B_x \\ B_y \\ B_z \end{pmatrix}_{x\leq 1.2} = 
\begin{pmatrix} 3.86859 \\ 0 \\ 0 \\ 0 \\ 167.345 \\ 0 \\ 2.1826182 \\ -2.1826182 \end{pmatrix}, \qquad 
\begin{pmatrix} \rho \\ v_x \\ v_y \\ v_z \\ p \\ B_x \\ B_y \\ B_z \end{pmatrix}_{x>1.2} = \begin{pmatrix} 1.0 \\ -11.2536 \\ 0 \\ 0 \\ 1.0 \\ 0 \\ 0.56418958 \\ 0.56418958 \end{pmatrix}.
\end{equation}
The density cloud, with density $\rho =10$ and radius $r = 0.15$, is centered at $(1.6,0.5)$ and considered to be in pressure equilibrium with the surrounding medium with $p = 1.0$.
Because of the strong gradients present in this problem, a lower CFL number, $C_{CFL}=1.75$ is used. The a-priori fallback approach (see section \ref{sec:apriori}) is also used with the threshold $\eta=3.6$. Less than $0.12\%$ of the reconstructions occur at lower order. The fraction of lower-order volume-to-area reconstructions, averaged on the whole simulation time, is as follows: 0.074$\%$ of a-priori second-order reconstruction, $5.10^{-5} \%$ of a-posteriori second-order reconstruction, $0.042\%$ of a-posteriori first-order reconstruction and $0.004\%$ of the area-average-to-point value transformations were switched off (see section \ref{sec:negcheck3}). Figure \ref{fig:cloudshock} contains color-coded contour plots of the density (top) and the total pressure (bottom) at time $t=0.1$ at resolution $800\times400$. The results are directly comparable to \cite{ROM15}. As can be observed in this figure, weak deviations from symmetry can naturally arise in situations with strong discontinuities due to higher-order numerics (cf. Figure 14 in \cite{ROM15}). We note here that these asymmetric fluctuations can nevertheless be strongly reduced by additional numerical diffusion, e.g., by lowering the threshold $\eta$. Moreover, we have also observed that at the limit $\eta \longrightarrow 0$, where only the TVD reconstruction is used, the results are as symmetric as the other tests presented in this paper.
\subsection{Numerical Dissipation}
\label{sec:dissipation}
In this section, we demonstrate the advantage of using higher-order schemes by comparing the effect of numerical dissipation for schemes of different orders of accuracy. Namely, the fourth-order scheme described in this paper is compared with a second-order scheme, which is the scheme that would be obtained if one would forcefully set the reconstruction order map everywhere to the value 0 (see section \ref{sec:apriori}). That is: the reconstruction procedure makes use of the TVD slope limiter, area$\leftrightarrow$point values transformations are deactivated, discontinuity propagation terms are removed, and a simple averaging of the four surrounding faces is taken for the edge-averaged electric field. Note that the magnetic cell interpolation (section \ref{sec:Bcelltoface}) and the time integration (section \ref{sec:timeINT}) still remain the same as the ones in the fourth-order scheme. 
\subsubsection{2D CPAW Test}
A 2D CPAW test is already discussed in the section \ref{sec:cpaw}.
Here we choose the amplitude $A = 7 \times 10^{-4}$. In figure \ref{fig:cpawdissi} we show the 1D cut of the $v_y$ profiles along the diagonal at resolutions $64^2$ and $128^2$ after $100$ periods, at time $t = 50$, where the red line represents the results from our fourth-order scheme, the blue line stands for the second-order scheme and the black line-points represents the initial conditions, taken as the reference solution. The solutions from our fourth-order scheme are very close to the reference solution exhibiting almost negligible numerical dissipation; however, a high dissipation in the second-order scheme is evident.
\subsubsection{MHD Vortex}
The MHD vortex problem is already discussed in the section \ref{sec:mhdvortex}. Here the resolution is chosen to be $128^2$ and a smaller domain, $[-5,5]\times[-5,5]$, is taken. In figure \ref{fig:mhdvortexdissi1} we show the color-coded contour plots of the magnetic energy after $10$ periods ($t=100$) from the second-order scheme (left) and fourth-order scheme (right). The fourth-order scheme seems to preserve the shape and strength of the vortex, whereas it gets deformed and weakens due to higher numerical dissipation in the second-order scheme. The numerical dissipation in the second-order scheme becomes more evident when we look at the temporal-evolution of the normalised magnetic energy as shown in figure \ref{fig:mhdvortexdissi2}.
\subsubsection{MHD Blast Wave}
This problem is also already discussed in the section \ref{sec:blastwave}. In figure \ref{fig:blastwavdissi}, we show the
1D cut of the density profiles along the diagonal of the computational domain at time $t=0.1$. The red and blue lines stand for the fourth and second-order schemes, respectively, obtained at resolution $128^2$. The black-line denotes the reference solution obtained from the second-order scheme at resolution $1280^2$. Here the left subplot covers the full domain along the diagonal, and the right subplot is zoomed in the range [0,0.7]. They depict that our fourth-order scheme is closer to the reference solution already at lower resolutions.
\section{Conclusion}
\label{sec:conclusion}
We have proposed in this paper a relatively simple fourth-order accurate finite volume CWENO scheme to study astrophysical MHD problems. This scheme uses area averages$\leftrightarrow$ point values transformations in order to maintain the accuracy of the scheme, allowing hence a time-efficient dimension-by-dimension approach where only one point-value is computed for each face. The fourth-order accuracy, shock capturing nature as well as the conservation of magnetic field solenoidality have been confirmed
through various multi-dimensional complex MHD problems. The extension of this method to even higher-order schemes should be
relatively straightforward, using higher-order 1D non-oscillatory reconstructions, higher-order area averages$\leftrightarrow$point values transformations and a higher-order time integrator.
\section*{Acknowledgments} 
PSV would like to acknowledge the support from CNRS and Aix-Marseille University, France in preparing this manuscript. 
JMT also gratefully acknowledges support by the Berlin International Graduate School in Model and Simulation based Research (BIMoS).  

\bibliographystyle{mnras}


\bsp	
\label{lastpage}
\end{document}